\documentclass[sigconf,authordraft,review=false]{acmart}

\AtBeginDocument{%
  }

\usepackage{bbm}
\usepackage{subcaption}
\usepackage{algorithm}
\usepackage{algpseudocode}
\newtheorem*{theorem*}{Theorem}
\newtheorem{theorem}{Theorem}
\usepackage{listings}
\usepackage{multirow}
\usepackage[normalem]{ulem}

\setcopyright{acmlicensed}
\acmJournal{PACMMOD}
\acmYear{2025} \acmVolume{3}\acmNumber{1 (SIGMOD)}  \acmArticle{69} \acmMonth{2}\acmDOI{10.1145/3709719}

\author{Kaiwen Chen}
\affiliation{%
\institution{University Of Toronto}
  \city{Toronto}
  \state{Ontario}
  \country{Canada}
}
\email{kckevinchen@cs.toronto.edu}
\orcid{0009-0007-6724-9870}

\author{Yueting Chen}
\affiliation{%
\institution{Seattle University}
  \city{Seattle}
  \state{Washington}
  \country{United States}
}
\orcid{0009-0003-6481-4011}
\email{ychen24@seattleu.edu}

\author{Nick Koudas}
\affiliation{%
\institution{University Of Toronto}
  \city{Toronto}
  \state{Ontario}
  \country{Canada}
}
\orcid{0000-0001-5648-0638}
\email{koudas@cs.toronto.edu}

\author{Xiaohui Yu}
\affiliation{%
\institution{York University}
  \city{Toronto}
  \state{Ontario}
  \country{Canada}
}
\orcid{0000-0001-8170-2327}
\email{xhyu@yorku.ca}

\orcid{0000-0001-5648-0638}
\email{koudas@cs.toronto.edu}

\begin{document}

\title{Reliable Text-to-SQL with Adaptive Abstention}



\begin{abstract}

Large language models (LLMs) have revolutionized natural language interfaces for databases, particularly in text-to-SQL conversion. However, current approaches often generate unreliable outputs when faced with ambiguity or insufficient context. 

We present Reliable Text-to-SQL (RTS), a novel framework that enhances query generation reliability by incorporating abstention and human-in-the-loop mechanisms. 
RTS focuses on the critical schema linking phase, which aims to identify the key database elements needed for generating SQL queries. It autonomously detects potential errors during the answer generation process and responds by either abstaining or engaging in user interaction. A vital component of RTS is the Branching Point Prediction (BPP) which utilizes statistical conformal techniques on the hidden layers of the LLM model for schema linking, providing probabilistic guarantees on schema linking accuracy. 

We validate our approach through comprehensive experiments on the BIRD benchmark, demonstrating significant improvements in robustness and reliability. Our findings highlight the potential of combining transparent-box LLMs with human-in-the-loop processes to create more robust natural language interfaces for databases. For the BIRD benchmark, our approach achieves near-perfect schema linking accuracy, autonomously involving a human when needed. Combined with query generation, we demonstrate that near-perfect schema linking and a small query generation model can almost match SOTA accuracy achieved with a model orders of magnitude larger than the one we use.

\end{abstract}



\received{July 2024}
\received[revised]{September 2024}
\received[accepted]{November 2024}
\newcommand{\OurMethod}{\textbf{RTS}}
\newcommand{\Uncertainty}{\textbf{uncertainty filter }}
\maketitle

\section{Introduction}









The advent of large language models (LLMs) has precipitated a paradigm shift in addressing canonical database challenges, encompassing data integration, information retrieval, and query comprehension. These models' sophisticated natural language understanding capabilities facilitate the extraction of structured data from unstructured text with unprecedented semantic fidelity. By leveraging transformer architectures and self-supervised learning on vast corpora, LLMs exhibit remarkable efficacy in discerning complex linguistic patterns and contextual nuances, thereby augmenting traditional database operations with enhanced semantic interpretation.
As an example, LLMs can seamlessly integrate disparate data sources \cite{chang2024survey, yang2024harnessing}, enhance the precision of information retrieval systems \cite{li2024llatrieval}, and improve the understanding and processing of complex queries \cite{brown2020language}.

The application of LLMs to text-to-SQL conversion \cite{gao2024text, guo2023prompting,hong2024knowledge,nan2023enhancing} has emerged as a focal point of research within the database community. 
Text-to-SQL encompasses the transformation of natural language queries into corresponding SQL statements, leveraging the underlying structure of relational databases. The significance of this research area lies in its potential to democratize data access, enabling non-technical users to interact with databases effectively. Text-to-SQL systems have wide-ranging implications for domains such as interactive data exploration, automated query generation, and natural language interfaces for database systems. However, the task presents several non-trivial challenges, including the inherent ambiguity of natural languages, the rigid syntactic and semantic constraints of SQL, and the necessity for precise schema mapping and entity resolution within the context of the query.

Fortunately, the advent of LLMs has yielded substantial improvements in accuracy and performance for text-to-SQL across various benchmarks,  
including surpassing traditional approaches by over 30\% in execution accuracy on the challenging BIRD \cite{li2024can} benchmark. 
This notable advancement has catalyzed a shift in research priorities within this evolving field, with current efforts now focusing on text-to-SQL prompt engineering \cite{arora2023adapt, gao2024text, tai2023exploring, nan2023enhancing}, fine-tuning LLMs using extensive SQL example repositories \cite{gao2024text, li2024codes} to improve SQL generation accuracy, and critically, improving schema linking \cite{pourreza2024din}.

\begin{figure}
    \centering
    \begin{subfigure}{0.45\textwidth}
        \includegraphics[width=\linewidth]{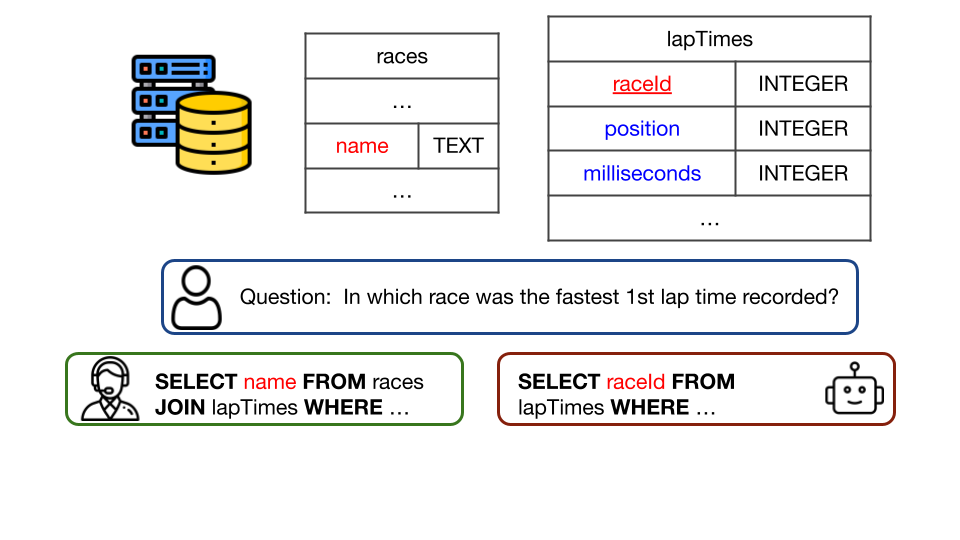}
        \vspace{-4em}
        \caption{Ambiguity in the query}
        \label{fig:ambiguous_text_to_sql}
    \end{subfigure}
    \hfill
    \begin{subfigure}{0.45\textwidth}
        \includegraphics[width=\linewidth]{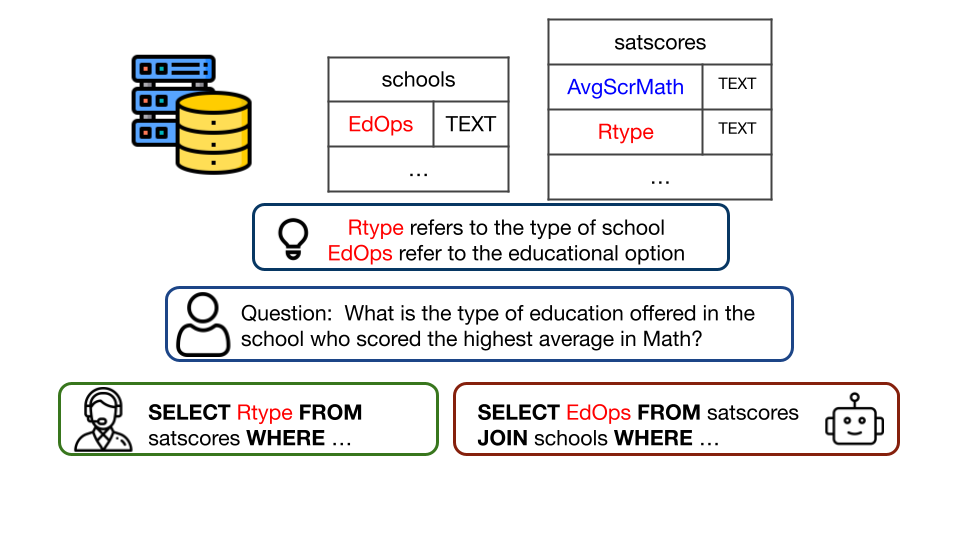}
        \vspace{-3em}
        \caption{Lack of metadata description at the schema}
        \label{fig:unclear_text_to_sql}
    \end{subfigure}
    \vspace{-1em}
    \caption{Example of Error-prone Text-to-SQL in BIRD Benchmark}
    \label{fig:error_text_to_sql}
    \vspace{-1em}
\end{figure}

A major challenge facing existing text-to-SQL approaches is the lack of ability to quantify and assess the confidence in the SQL queries generated, as they predominantly utilize LLMs as opaque
processing units. Crucially, 
these systems are designed to generate outputs to every input they receive, even when they may not be adequately equipped to do so, due to the lack of domain-specific knowledge or additional contextual information.
This often leads to outputs that are essentially educated guesses, which can compromise the accuracy and relevance of the results.

To elucidate this issue, we present a case study from the BIRD benchmark, illustrated in Figure \ref{fig:error_text_to_sql}(a). In this example, the user query seeks to identify the race with the minimum first lap time. The term ``race'' introduces semantic ambiguity, as it could reference either the race name in the races relation or the raceId in the lapTimes relation. Consequently, while the ground-truth SQL and the model-generated query may yield divergent result sets, both should be considered valid interpretations given the inherent ambiguity in the natural language query. Similarly, consider the scenarios illustrated in Figure \ref{fig:error_text_to_sql}(b), where the language model makes an erroneous generation not due to semantic ambiguity but simply because the schema does not provide enough information to aid correct generation. In this example, it is unclear whether {\em EdOps} or {\em Rtype} stands for the {\em type of education}, even with the help of the database description.


The issues illustrated by the above examples are especially true in the schema linking phase of query generation.
Schema linking involves mapping the natural language components of the query to the corresponding elements in the database schema. This phase is particularly error-prone in text-to-SQL pipelines \cite{pourreza2024din} due to the complexity of accurately resolving entities and aligning them with the correct schema elements,   
and it has been the primary focus of recent advancements in the field \cite{chang2020zero, zhang2019editing, li2023graphix,hui2022s2sql,cao2021lgesql}.

\begin{figure}[]
    \centering
    \includegraphics[width=\linewidth]{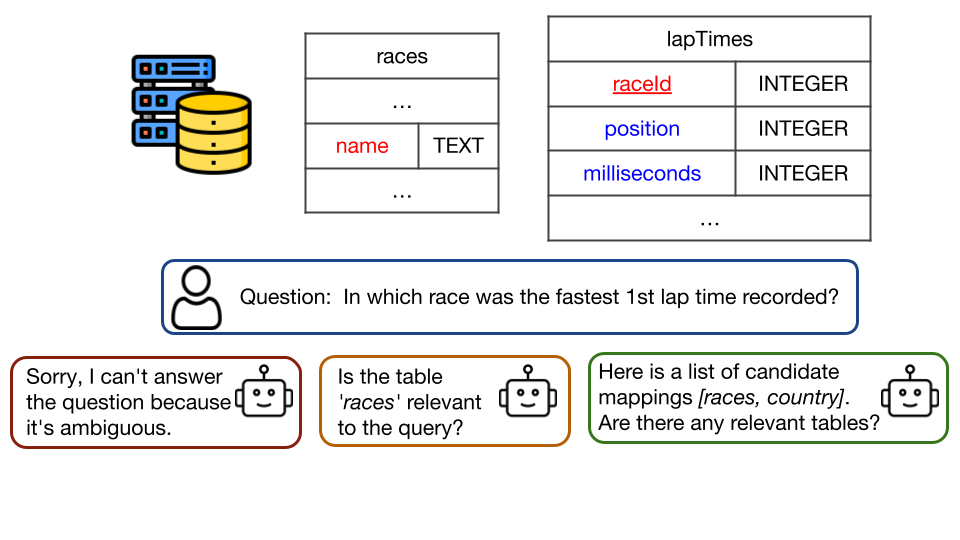} 
    \vspace{-2em}
    \caption{Example of the \OurMethod ~Framework: When probable errors are detected during answer generation, the model will either abstain from generating the query (left), ask the user for confirmation of the potential schema (middle), or prompt the user for hints (right).}
    \label{fig:framework-text-to-sql}
    \vspace{-1em}
\end{figure}

In light of these challenges, we introduce a novel framework called \textbf{R}eliable \textbf{T}ext-to-\textbf{S}QL (\OurMethod) that enhances query generation reliability. 
{ The defining feature of our proposal is the ability to abstain during the answer-generation process, specifically designed to identify uncertainties and allow the user to resolve ambiguities by choosing the interpretation that best aligns with their intent}.
 Our focus is on the schema linking phase as it is the most error-prone and crucial for the overall accuracy of text-to-SQL systems. 
As illustrated in Figure \ref{fig:framework-text-to-sql}, our framework can autonomously detect errors during schema linking and react independently. For example, as shown in the leftmost chatbox, when the model detects a probable error during the generation of the answer, we may opt to halt the automatic mapping process to avoid likely erroneous linking. Alternatively, we could engage in an interactive refinement process. This process involves presenting the human user with the targeted disambiguation question (middle chatbox) for confirmation. Alternatively, a list of probable mapping candidates (rightmost chatbox) can be presented to the human user or an expert LLM for selection. { Our approach leverages language models and integrates a learned ambiguity detection mechanism during inference. Moreover, by employing conformal prediction techniques \cite{papadopoulos2002inductive, vovk2005algorithmic,barber2023conformal}, we provide statistical guarantees on the accuracy of the data generation process during schema linking, thereby enhancing the robustness and reliability of the entire query generation process.}

We validate the effectiveness of our proposed framework through a series of experiments on the BIRD benchmark \cite{li2024can}. Our results underscore the potential of combining transparent-box LLMs with human-in-the-loop processes to create more robust and industry-ready solutions for natural language interfaces to databases.

To summarize in this paper we make the following contributions:
\begin{enumerate}
    \item We conduct a comprehensive investigation into the reliability of Text-to-SQL generation, identifying key challenges and opportunities for improvement.
    \item We propose a novel framework, \OurMethod  ~which automatically detects uncertainty during the schema linking process and abstains from generating potentially erroneous queries. \OurMethod ~can be adopted as an addon, by any transparent-box LLM responsible for schema linking.
    \item We theoretically analyze important parts of the framework establishing probabilistic guarantees on its overall accuracy.
      \item \OurMethod~ introduces techniques to instigate human intervention when required allowing the model to employ auto-correction based on expert feedback.
    \item  We evaluate our proposed framework through a series of rigorous experiments, utilizing benchmark datasets. Overall we achieve near-perfect schema linking accuracy even on the challenging BIRD benchmark utilizing \OurMethod ~with human intervention when required. This has strong accuracy implications during subsequent SQL generation.
\end{enumerate}

This paper is organized as follows: Section \ref{sect:background} discusses background material and observations that are relevant to the remainder of the paper. Section \ref{sect:rts} presents the \OurMethod ~framework and its full specification. In Section \ref{sect:exp} we present the results of a detailed experimental evaluation, followed by related work and conclusions in Sections \ref{sect:related} and \ref{sect:conclude}.

\section{Background and Problem Statement} \label{sect:background}

\subsection{General Text-to-SQL Framework}
Text-to-SQL translates a natural language question $\mathbf{Q}$ into an SQL query for database $\mathbf{D}$. 
We outline steps and design space of LLM-powered Text-to-SQL frameworks.
\paragraph{\textbf{Pre-Processing:}}
Pre-processing, links schema to the question, aiding SQL generation. Schema linking is crucial but error-prone in LLM-based systems \cite{pourreza2024din}.
\paragraph{\textbf{SQL Generation:}}
SQL generation uses LLMs to translate questions into queries. Models take linked schema and questions as input, outputting SQL \cite{pourreza2024din, ren2024purple, gao2024text}. New, better-performing models constantly emerge.
Prompt engineering enhances fine-tuned models via zero-shot \cite{chang2023prompt} or few-shot learning \cite{gao2024text}. Multi-step generation \cite{xie2024decomposition, pourreza2024din, gao2024text, nan2023enhancing, ren2024purple}, like chain-of-thought reasoning \cite{tai2023exploring}, improves accuracy.
\paragraph{\textbf{Post-Processing:}}
Post-processing refines generated SQL. Common strategies include:
\begin{enumerate}
\item Self-Correction: Models review generated SQL to address issues \cite{pourreza2024din}.
\item Self-Consistency: Multiple valid queries are executed; voting determines the most consistent \cite{nan2023enhancing}.
\item Execution-Guided SQL Selector: Sequentially executes generated SQLs, choosing the first error-free execution \cite{ni2023lever, guo2023retrieval}.
\end{enumerate}

\subsection{Schema Linking}
Schema linking involves precisely identifying all the relevant columns and tables in a large database that are necessary to answer a given natural language query. At a high level, in LLM-based methods, this process involves querying an LLM possibly after supervised fine-tuning\footnote{In the literature there are numerous approaches to do this \cite{li2024codes}.}. This LLM (referred to as the {\em schema linking model}) accepts a natural language query as input (along with the database schema and other applicable metadata) and outputs the tables and columns relevant to the query. 


Specifically, given a database $\mathbf{D}$ with tables $\mathcal{T} = \{T_1, T_2, \ldots, T_{\nu}\}$, the columns $\textbf{c}_{i}  = \{c_{i1}, c_{i2}, \ldots, c_{i\kappa}\} $ for each table $T_i$, and a natural language question $\mathbf{Q}$, the goal of schema linking is to:


\begin{enumerate}
    \item Identify the relevant tables 
    $\mathcal{T}'\subseteq \mathcal{T}$ for forming an answer to $\mathbf{Q}$ (referred as {\em table linking})
    \item Extract the relevant columns $\textbf{c}'_i$ from $\textbf{c}_{i}$ for each table $T_i \in \mathcal{T}'$ for forming an answer to $\mathbf{Q}$ (referred as {\em column linking})
\end{enumerate}
These lists of tables $\mathcal{T}'$ and columns $\textbf{c}'$ will be used to formulate an SQL query that answers the question $\mathbf{Q}$.

\begin{table}[h]
\centering
\caption{Text-to-SQL Performance with Chess\cite{talaei2024} on Bird Development Set}
\begin{small}
\begin{tabular}{@{}lc@{}}
\toprule
Schema Linking Configuration               & Execution Accuracy (EX) \\ \midrule
Correct tables + Correct columns     & 72.4 \\
Full  tables + Full columns                  & 64.52 \\ \bottomrule
Best reported based method~\cite{pourreza2024chasesqlmultipathreasoningpreference} & 73.01 \\ \bottomrule
\end{tabular}
\end{small}
\label{tab:schema_linking_motivation}
\end{table}

Accurate schema linking is crucial for enhancing the downstream construction of complex SQL queries. 
To demonstrate the significant impact of schema linking on the accuracy of generating complex SQL queries, we conduct experiments that involve supervised fine-tuning of the small-scale language model Deepseek-7B \cite{deepseek-llm} on the BIRD training dataset \cite{li2024can}, using various schema configurations:
(1) only the relevant tables and columns are provided (Correct tables + Correct columns), (2) the relevant tables are provided but might include irrelevant columns (Correct tables + Full columns), and (3) the entire database schema is used (Full tables + Full columns). We measure the effectiveness of each configuration based on execution accuracy, which is determined by comparing the execution results of both the predicted queries (the queries provided by the text-to-SQL model) and the gold queries (ground-truth annotated SQL queries) against the BIRD database.

As demonstrated in Table \ref{tab:schema_linking_motivation}, even a small-scale language model(~34B), when coupled with accurate schema linking, achieves comparable performance to the best methods based on large-scale language models like Gemini (referred to as {\em Best reported based method} in Table~\ref{tab:schema_linking_motivation}) as reported on the BIRD leader board \cite{birdwebsite}. 



Perfect schema linking is challenging due to:
\begin{itemize} 
\item Semantic Complexity: Schemas contain nuanced relationships and naming conventions, requiring deep comprehension across domains.
\item External Knowledge Integration: Linking often needs domain-specific information beyond the schema, which can be difficult to acquire and integrate.
\item Language Ambiguity: Queries may contain ambiguous terms, requiring contextual understanding to resolve accurately.
\end{itemize}

Schema linking complexities demand effort in optimization, curation, and customization, especially in new environments.
We propose an error-aware abstention mechanism to enhance model reliability. This allows flagging potential errors, mitigating erroneous mappings and improving pipeline trustworthiness.
Integrating abstention is crucial for adoption and reliability. It reduces engineering overhead in new domains and establishes a feedback loop for continuous learning, facilitating integration into existing workflows.

\subsection{Text-to-SQL with Abstention}


Using a schema linking model, we aim to improve token generation during the answer formation process - specifically in predicting tables and columns pertinent to the query - by identifying critical {\em branching points}.
These branching points are those generated tokens that are erroneous and steer the generation away from the correct answer (alternatively instigate {\em branching off} from the correct answer). Moreover, we will establish guarantees on the probability of detecting a branching point during a generation. Detecting branching points forms the basis of our abstention mechanisms.

Note that we allow the schema linking model to abstain during the pre-processing step, which has two major advantages:
\begin{enumerate}
\item \textbf{Computation Efficiency}: 
Enabling the model to abstain during the pre-processing phase significantly reduces unnecessary computations, which are most likely to lead to erroneous SQL queries, thereby enhancing efficiency during SQL generation. 
\item \textbf{Seamless Integration with Advanced Models}: Our approach facilitates the seamless integration of newly developed text-to-SQL models in the SQL generation phase without extensive re-engineering. We view this as highly important, as SQL generation models are advancing rapidly with new models available frequently. As a design principle, our proposal is agnostic to the SQL generation model, and we can adopt the best performing one. Abstention during the schema linking process ensures that only high-confidence schemas are delivered to the SQL generation models, thereby improving overall performance and reducing errors.
\end{enumerate}

The abstention mechanism also safeguards against low-confidence predictions, enhancing the overall reliability of the system. 
When the model abstains, as we will detail below, alternative strategies such as human-in-the-loop can be employed to ensure the computation progresses with accuracy. This involves a human expert reviewing the model output and possible suggestions, and resolving the issue that instigated abstention, thereby allowing the model to continue generation while maintaining the system reliability.
\section{Reliable Text-to-SQL} \label{sect:rts}

In this section, we present our proposed framework, Reliable Text-to-SQL (\OurMethod), detailing all of its components.

\begin{figure}[ht]
    \centering
    \begin{subfigure}[b]{0.23\textwidth}
        \centering
        \includegraphics[width=\textwidth]{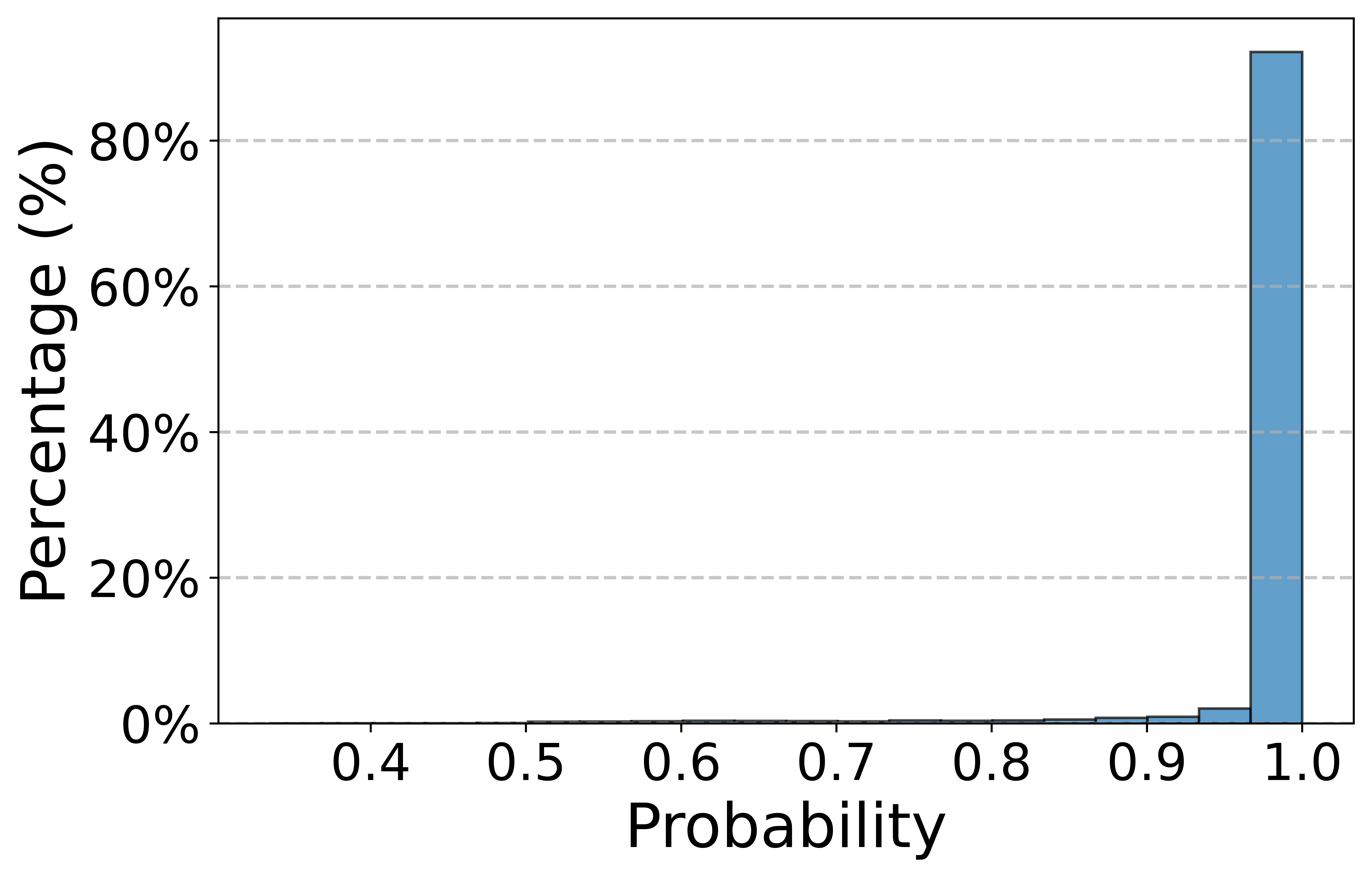}
        \caption{Probability distribution for the next token generation}
        \label{fig:fig2}
    \end{subfigure}
    \hfill
    \begin{subfigure}[b]{0.23\textwidth}
        \centering
        \includegraphics[width=\textwidth]{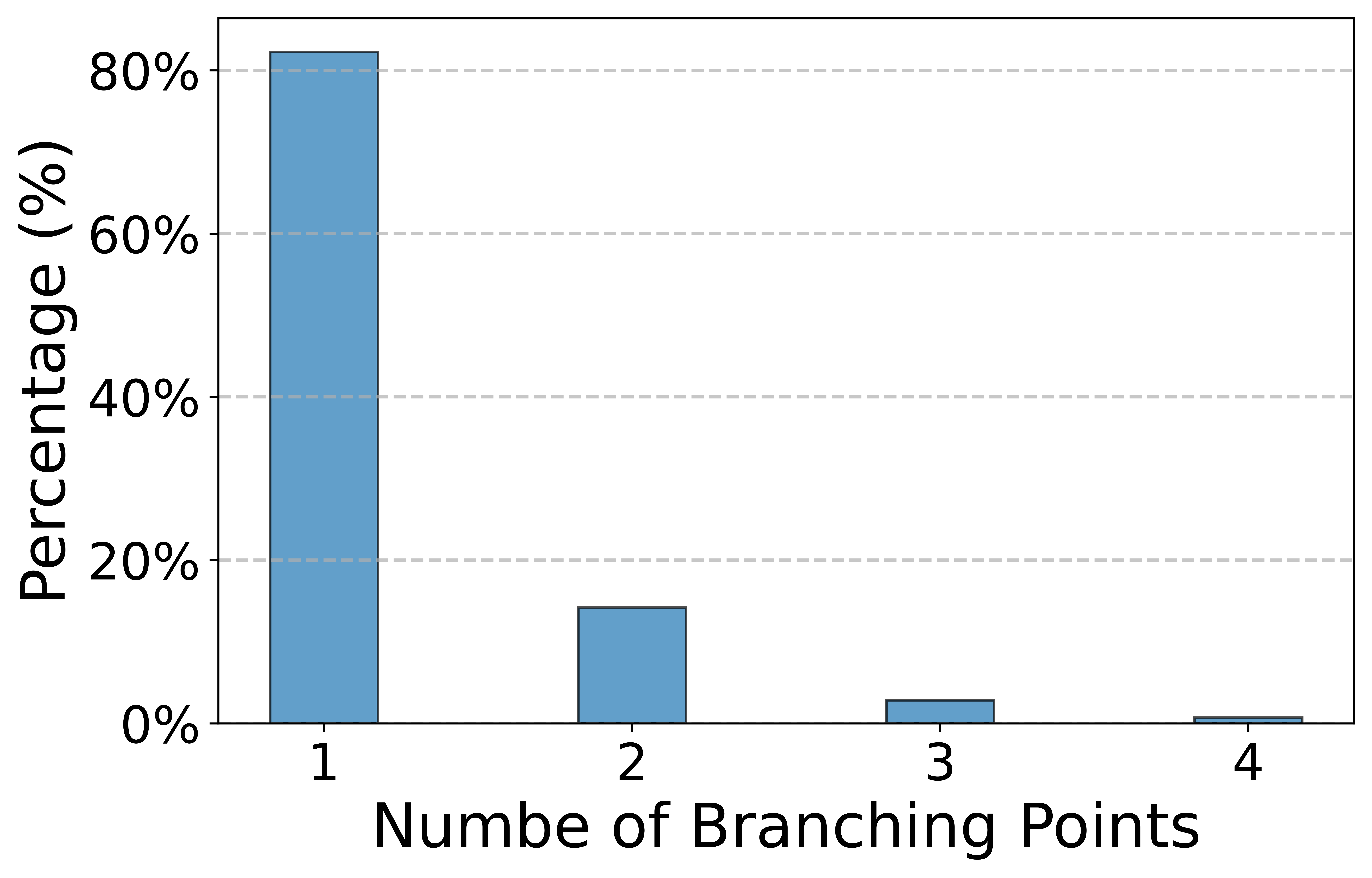}
        \caption{Number of branching points during inference}
        \label{fig:dist_branch}
    \end{subfigure}
    \vspace{-1em}
    \caption{Statistic For Finetuned Deepseek Model on the BIRD Development Dataset}
    \label{fig:side_by_side}
\end{figure}


\subsection{Branching Points During Schema Linking}

In this section, we outline our strategy for developing an abstention mechanism in the Text-to-SQL schema linking process. In the literature,  quantifying
uncertainty in LLM predictions is an active research topic but is recognized as challenging due to the nuances of language semantics and form \cite{geng2024survey}. Recent studies have proposed methods to address these challenges for free-form language generation in LLMs \cite{lingenerating, geng2024survey}. These methods leverage entropy \cite{vazhentsev2023efficient}, semantic analysis \cite{kuhn2023semantic}, and logits or hidden state information \cite{azaria2023internal, li2023inference, ren2023out} to quantify uncertainty in pre-trained LLMs during 
generation. 

Our problem of interest differs from free-form language generation in two key aspects. First, our generation task is semi-structured, expecting the model to output only a list of tables and columns present in the given schema. Second, and most importantly, our focus is on a supervised fine-tuned LLM for schema linking. As previously reported, supervised fine-tuned LLMs exhibit over-confidence \cite{xiong2024can} in their predictions. Namely, during answer generation, the probability distribution of the next generated token is highly skewed, regardless of the correctness of the generation. To demonstrate this, we recorded the softmax probability during schema-linking generation using a supervised fine-tuned Deepseek model on the BIRD development dataset. As illustrated in Figure \ref{fig:fig2},  the output probabilities are concentrated around 1 for both correct and incorrect schema predictions.
This phenomenon renders useless any logit-based method built on the intuitive expectation that an error in the generation is likely to correspond to a token generated with low probability values \cite{vazhentsev2023efficient}. 

For these reasons, we develop techniques to identify erroneous token generations tailored to the schema linking problem during Text-to-SQL. To achieve this, we begin by examining the errors made during the schema-linking process (assuming access to ground truth data) which is critical for accurate Text-to-SQL generation. In this section, we will use table linking as an illustrative example, as the methodology for column linking is similar.

Let $\mathcal{T}$ be the set of tables we have to schema link against. Let $\mathcal{T}^{t}$ be the set of tokens in these tables. We constrain the model's token level generation to only generate tokens in $\mathcal{T}^{t}$ utilizing constraint generation \cite{beurer2024guiding}.
We denote the model's token-level generation for a specific schema linking instance, as $\hat{x}_1, \dots, \hat{x}_m$, while the ground truth tokens are $x_1, \dots, x_m$. Each generated token $\hat{x}_i$ is compared with the corresponding ground truth token $x_i$, and the first token where the model generation deviates from ground truth is referred to as the \emph{branching point}, denoted as $\hat{x}_b$ for the generated token and $x_b$ for the ground truth token. Formally, the branching point satisfies the following conditions: $\hat{x}_b \neq x_b$, and for all $i < b$, $\hat{x}_i = x_i$. In other words, the branching point indicates the position where the model generation begins to diverge from the correct sequence, marking the first mistake made by the model.

\begin{figure}
    \centering
    \includegraphics[width=\linewidth]{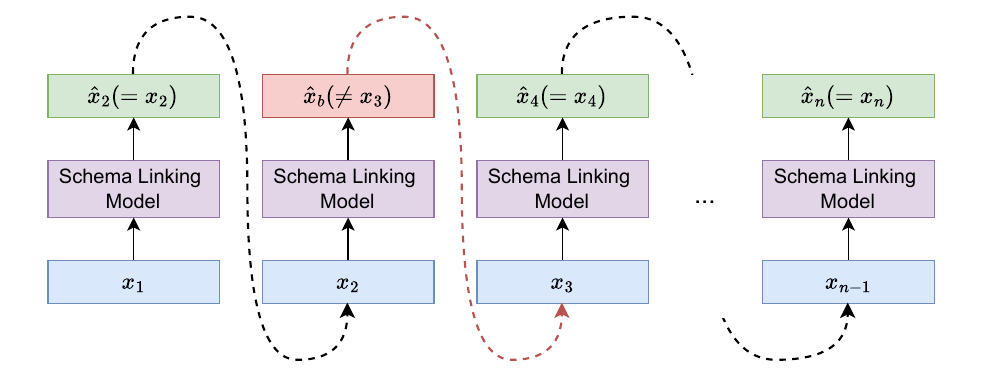}
    \vspace{-1em}
    \caption{Branching Points: During the generation of schema linking tokens, we compare the predicted token with ground truth. If the prediction diverges from ground truth, as in the prediction $\hat{x_3}$ above, teacher forcing will provide the correct token $x_3$ to the input to continue generation. }
    \label{fig:branching_point}
\end{figure}

Further, at the branching point, we employ \textit{teacher forcing} \cite{williams1989learning} which replaces the incorrect token with the corresponding ground truth token and allows the model to continue generating, as demonstrated by the red arrow in Figure \ref{fig:branching_point}. This process is repeated until the entire generation of the model matches the ground truth. It should be noted that during each teacher forcing step, we manually intervene to correct the mistake made by the model and continue the generation until (possibly) we encounter a new branching point using the same criteria. Hence, it is possible that multiple branching points may be identified during a single answer generation. (i.e. The model mistakenly deviates from the correct generation after correcting the previous mistake, one or more times.)

To highlight this, we conducted the above experiment on the BIRD development dataset with a supervised fine-tuned DeepSeek model on the corresponding training data.
We record the number of branching points within each {\em erroneous generation}, as shown in Figure \ref{fig:dist_branch}. In this experiment, we make two important observations:
\begin{enumerate}
    \item The number of branching points during erroneous schema linking is small for each schema linking operation, with more than 90\% of the erroneously generated sequences containing only one or two branching points.
    \item If these branching points can be detected and fixed, the model generates the correct schema linking result.
\end{enumerate}

Based on this observation, if one can accurately and effectively identify branching points during answer generation, such identification can be used as a basis of an abstention mechanism. Moreover, since the number of branching points in each generation is small, it is possible to progressively correct the model output by human intervention. Therefore, we propose to use branching points as a measure of uncertainty during schema linking generation. Specifically, the uncertainty over generating the next token $x_i$ is represented by the probability that the generated token is a branching point.

We build our {\em branching point predictor} on top of each hidden state.  We construct one classifier per hidden state
and utilize the results during each token generation to quantify the probability of that token being a branching point.
Specifically, we observe $\mathcal{N}$ schema linking queries posed to the schema linking model containing both correct as well as erroneous generations. For each schema linking query we assume the availability of the correct schema linking output (ground truth). We trace each answer generation one token at a time and construct the dataset 
\begin{align*}
d_l &= \{(h^i_1,\dots h^i_n), (s^i_1, \dots ,s^i_n)\}_{i=1}^m \\
\mathcal{D}_{branch} &= \{d_l\}_{l=1}^\mathcal{N}
\end{align*}
where $d_l$ represents a single generation process (i.e., the answer of the LLM for a specific schema linking task for a given query) involving $m$ tokens in a network with $n$ layers. Within this generation process, $h^i_j$ is the $j$-th hidden state vector for the $i$-th token, and $s^i_1 = \dots = s^i_n $ is a set of boolean variables indicating the ground truth value of the corresponding token (whether the $i$-th token corresponds to a branching point, assuming values 1 if it is and 0 otherwise). $\mathcal{D}_{branch}$ is the set of all $d_l$ over $\mathcal{N}$ schema linking queries.

\begin{figure}
    \centering
    \includegraphics[width=\linewidth]{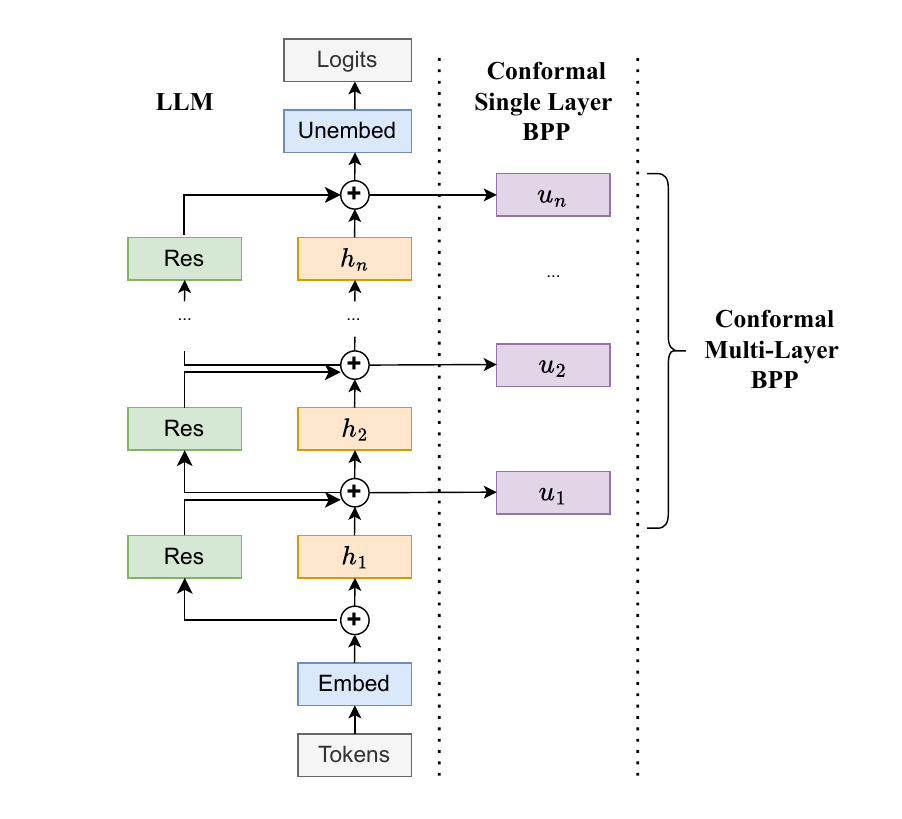}
    \vspace{-2em}
    \caption{Branching Point Predictor: Calibrated Classifiers per layer are trained to predict branching points. We aggregate the predictions in a principled manner to produce a final prediction.}
    \label{fig:transformer}
\end{figure}

$\mathcal{D}_{branch}$ is used to train $n$ classifiers, one for each hidden layer, denoted as $u_1 \dots u_n$, each consisting of a two-layer perceptron (MLP) classifier. 
During a brand new schema linking query, a set of hidden
state vectors (one per layer) will be produced for each token during inference.
The classifiers $u_1 \dots u_n$, when they are provided with the hidden state vector corresponding to their layer will output scores $\hat{s}_1, \dots \hat{s}_n$ for the corresponding token. The score $\hat{s}_i, 1 \leq i \leq n$ constitutes a prediction at each layer signifying whether the specific token been generated is a branching point, according to the classifier $u_i$. We present the high-level architecture in Figure \ref{fig:transformer}.

\subsection{Branching Point Predictor ({\textbf{BPP}})}

\label{sec:bppsec}

{ We now present the design of our branching point predictor (BPP) that utilizes the predictors $u_1, \dots, u_n$ encompassing probabilistic guarantees for the detection of branching points during schema linking token generation. We start by introducing a formalism of conformal prediction \cite{vovk2005algorithmic, papadopoulos2002inductive} \footnote { Due to space constraints, we only provide an overview of the main technique. Full details are available elsewhere \cite{shafer2008tutorial,angelopoulos2022gentleintroductionconformalprediction,barber2023conformalpredictionexchangeability,farinhas2024nonexchangeableconformalriskcontrol}.} that will be utilized in the following sections.}

\subsubsection{Conformal Prediction}
Let $(x, y)$ be a random pair with joint distribution $D$ on $\mathcal{X} \times \mathcal{Y}$, where $\mathcal{X}$ is the feature space and $\mathcal{Y}$ is the label space. For a given 
\textit{error level} $\alpha \in (0, 1)$, for any classifier, conformal prediction constructs a \textit{prediction set} $C : \mathcal{X} \to 2^\mathcal{Y}$ such that \cite{vovk2005algorithmic, papadopoulos2002inductive}:
\begin{equation}
\label{eq:conf}
p(y \in C(x)) \geq 1 - \alpha
\end{equation}
Note that $C(x)$ is a subset of the label space and conformal prediction bounds the probability that the correct label is a member of that set.
This guarantee is valid under the assumption of data exchangeability, which is a weaker condition than the independent and identically distributed (i.i.d.) assumption \cite{vovk2005algorithmic}. It remains applicable, albeit in a modified form, even when this assumption is not met \cite{barber2023conformal}, thereby extending its generality and applicability.
Let $(x_1, y_1), ..., (x_{N_d}, y_{N_d})$ be the held-out calibration data, and $(x_{test}, y_{test})$ be a test point. Conformal prediction proceeds as follows:
{ First, define a nonconformity measure $A : (\mathcal{X} \times \mathcal{Y}  \times (\mathcal{X} \times \mathcal{Y})^{N_d} \to \mathbb{R})$. The intent of $A$ is to quantify how related/unrelated a new point is to the calibration data. 
Then, for each $y \in \mathcal{Y}$, we can compute its nonconformity score $R_i$ as follows:
$$R_i = A((x_i, y_i), {(x_j, y_j) : j \neq i}), \text{ for } i = 1, ..., {N_d}$$
and the nonconformity score for the test data point  as:
$$R_{test}^{y} = A\left((x_{test}, y), {(x_i, y_i) : i = 1, ..., {N_d}}\right)$$
We utilize the nonconformity score to compute the $\pi$-value for each $y$:
$$\pi(y) = \frac{|{i : R_i \geq R_{test}^{y}}| + 1}{{N_d} + 1}$$
Finally, the prediction set is then defined as:
$$C(x_{test}) = \left\{y \in \mathcal{Y} : \pi(y) > \alpha\right\}$$}

Under the assumption that $\mathcal{X} \times \mathcal{Y}$ is exchangeable, the prediction set satisfies Equation \ref{eq:conf}. 
It is important to highlight that this form of guarantee is marginal \cite{vovk2005algorithmic}, indicating that the probability is calculated over a random draw from the calibration data and the test point. A similar guarantee can be provided for non-exchangeable distributions \cite{barber2023conformal}.

\subsubsection{Conformal Single Layer BPP (sBPP)}
\label{sec:single}

We propose to employ the conformal prediction methodology to construct prediction sets for a classifier $u_i$. This classifier assigns scores to tokens generated at layer $i$ of the schema linking model, indicating the probability (a score) of each token being a branching point, as depicted in Figure~\ref{fig:transformer}.

The classifier $u_i$ will be trained using data from $\mathcal{D}_{branch}$, specifically the $i$-th hidden state  
vectors and all tokens from schema linking tasks $d_l \in \mathcal{D}_{branch}$, along with their corresponding ground truth values ($h_i^{j}, s_i^{j}, 1 \leq j \leq m$, $\forall l, d_l \in \mathcal{D}_{branch}$). Let $D^{i}$  denote this dataset, and $D_{c}^{i}$ be a subset of $D^{i}$ designated as the calibration set. The training of $u_i$ will utilize the data in $D^{i} \setminus D_{c}^{i}$.

Each observation in the calibration set comprises a feature vector $x_i \in \mathcal{X}$ (i.e., the hidden state vector $h_i$ for a given token) and its associated ground truth label $y_i \in \mathcal{Y}$ (i.e., the value $s_i$ for that token, indicating whether it is a branching point).

Utilizing this calibration dataset, we define the non-conformity score of a data point $x$ as $1 - p(y^*\mid x)$, where $p(y^*\mid x)$ is the softmax probability of the true class $y^*$ for classifier $u_i$. This score quantifies the degree of deviation or non-conformity of a new observation from the training data distribution. Note that the non-conformity score increases when the classifier $u_i$ makes an incorrect prediction (as the evaluation takes place utilizing ground truth data at this stage).

We then establish a threshold $\epsilon$ as the $\lceil(|D_{c}^{i}| + 1)(1 - \alpha)/|D_{c}^{i}| \rceil$-th quantile of the non-conformity scores computed over the calibration set, where $|D_c^i|$ denotes the cardinality of the calibration set. For a new test point $x_\text{test}$, we generate the prediction set as:
$$C(x_{test}) = \left\{y \mid p(y\mid x_{test}) \geq 1 - \epsilon\right\}$$
Under the assumption of exchangeability for the calibration dataset $D_{c}^{i}$, we can guarantee that:
$$p(y^* \in C(x_{test})) \geq 1- \alpha$$
where $y^*$ represents the true label for $x_{test}$. 
This approach allows us to construct prediction sets with a specified 
error level,
providing a foundation for the detection of branching points.

In the general case, one cannot assume exchangeability, and non-exchangeable conformal prediction can be utilized \cite{barber2023conformal}. The framework is similar, but the way the calibration set is processed and the determination of the threshold $\epsilon$ are adjusted to account for the differences between the distributions of the test and calibration data. 

{Specifically, we transform $D_{c}^{i}$ by associating each hidden state vector $h_i \in D_{c}^{i}$ with its non-conformity score $\sigma_i = 1-p(y | h_i)$ computed from $u_i$. Let $D_{c}^{i'}$ be the transformed calibration set of pairs $(h_i,\sigma_i)$. At test time, given a test hidden state $h^{*}$
we compute its $K$ nearest neighbors in $D_{c}^{i'}$ and compute the weights $w_k = \exp(-||h^{*}-h_k||_{2}^{2}/\tau)$ for a hyper-parameter $\tau$ and $1 \leq k \leq K$.}
Essentially we assess the "influence" each calibration point has on the test point at hand.
After normalization $\hat{w_i} = w_i/(1+\sum_{i=1}^{K}w_i)$
we compute a new threshold $\hat{\epsilon} = \inf\{\epsilon|\sum_{i=1}^{K}\hat{w_i}\mathbf{1}\{\sigma_i < \epsilon\} \geq 1-\alpha\}$, where $\mathbf1\{.\}$ is an indicator function. 
The estimation proceeds as in the exchangeable case; the bound of Equation \ref{eq:conf} is slightly looser in the non-exchangeable case, incorporating correction terms \cite{barber2023conformal}. Full details and associated analysis are available elsewhere \cite{barber2023conformal}.


\subsubsection{Conformal Multi-Layer BPP (mBPP)}

We repeat the construction in Section~\ref{sec:single} for each of the $n$ classifiers $u_1 \ldots u_n$. Let $C_i = C_i(x) \subseteq \{0,1\}$ denote the conformal prediction set derived by $u_i$ on some input $x$. This set includes the labels (0, 1, or both) predicted by classifier $u_i$ for the corresponding token. To ease notation, we will assume that for the calibration set of each $u_i$ exchangeability holds, but all discussion applies equally for the non-exchangeable case.

{\paragraph{Majority Vote}

We first present an intuitive and robust procedure for aggregating the predictions of the classifiers to declare whether the test token is a branching point, based on majority voting. 
Specifically, let $C_1, C_2, \dots C_n$ denote the prediction set generated by the $n$ branching point predictors $u_i$, each based on a different hidden layer in the schema linking model.} Given a threshold $\theta \in [0,1)$, we define $C^{\theta}$ to include all predictions (0 or 1) that 
appear in at least $\theta$ fraction of the
prediction sets,
$$ C^{\theta} = \left\{ c \in \{0,1\}: \frac{1}{n}\sum_{i=1}^{n}\mathbf{1} \{ c \in C_i \}  > \theta \right\},$$
which has the following important property:
\begin{theorem}
\label{th:1}
    Let $C_1, \ldots, C_n$ be prediction sets with properties as per Equation \ref{eq:conf} and  $c^*$ denote the correct label. Then,
    $$p(c^* \in C^{\theta}) \geq 1-\alpha/(1-\theta)$$
\end{theorem}
\begin{small}
\begin{proof}
    $$p(c^* \notin C^{\theta}) = p\left(\frac{1}{n}\sum_{i=1}^{n}\mathbf{1} \{ c^* \notin C_i \} \geq 1-\theta\right) $$
    Noted that $\mathbf{1} \{ c^* \notin C_i \}$ is a Bernoulli random variable (referred to as $\phi_i$), with its expectation $E(\phi_i)\leq \alpha$.
    Therefore, by Markov’s inequality,
    \begin{align*}
        p(c^* \notin C^{\theta}) &= p\left(\frac{1}{n}\sum_{i=1}^{n}  \phi_i  \geq 1-\theta\right) \\
        &\leq \frac{E(\frac{1}{n}\sum_{i=1}^{n}\phi_i)}{1-\theta}
        = \frac{\frac{1}{n}\sum_{i=1}^{n}E(\phi_i)}{1-\theta} \\
        &\leq \frac{\alpha}{1-\theta}
    \end{align*}
Thus, $$p(c^* \in C^{\theta}) = 1-p(c^* \notin C^{\theta}) \geq  1-\alpha/(1-\theta)$$
\end{proof}
\end{small}

Theorem \ref{th:1} makes no assumptions on how the prediction sets $C_i$ are related. For the special case of $\theta = 1/2$, it becomes the majority vote and specifies that by aggregating the predictions for the labels, the majority vote provides a marginal probability guarantee of at least $1-2\alpha$.  The actual guarantee depends on the properties of sets $C_i$. For example, if the sets $C_i$ are independent, for $\theta=1/2$ one can show using Hoeffing's inequality that the probability of a false negative is at most $\exp^{(-2n(1/2-\alpha)^{2})}$, which goes to 0 fast as $n$ increases (the number of layers of the networks and thus the corresponding number of classifiers increases). Alternatively, if all the sets are the same, the probability of a false negative is upper-bounded by $\alpha$. The majority vote provides a balance between false negatives (missing a branching point) and false positives (erroneously declaring a point as branching when it is not). In our application, false negatives are harmful as they will lead to erroneous schema linking. False positives, however, when present, are not harmful in the sense of providing wrong schema linking information. Instead, will trigger abstention or human intervention (see Section \ref{sec:mit}) but they need to be minimized as well.

We next show that the size of $C^{\theta}$ in the worse-case, can be bounded:
\begin{theorem}
    Let $|C_1|, \ldots, |C_n|$ denote the size of each prediction set. Then,
    $$|C^{\theta}| \leq \frac{1}{n\cdot \theta} \sum^{n}_{i=1}|C_i|$$
\end{theorem}
\begin{small}
\begin{proof}
\begin{align*}
|C^{\theta}| &= \sum_{c} \mathbf{1} \left\{\frac{1}{n}\sum_{i=1}^{n}\mathbf{1} \{ c \in C_i \} \geq \theta\right\} 
\leq \sum_{c} \frac{1}{n\cdot \theta}\sum_{i=1}^{n}\mathbf{1} \{ c \in C_i \} \\
&= \frac{1}{n\cdot \theta}\sum_{i=1}^{n} \left(\sum_{c} \mathbf{1} \{ c \in C_i \} \right)
= \frac{1}{n\cdot \theta}\sum_{i=1}^{n} |C_i|\\
\end{align*}
where the first inequality is based on the fact $\mathbf{1} \{x \geq 1\} \leq x$.
\end{proof}
\end{small}

Notice that for $\theta = \frac{1}{2}$,  $|C^{\theta}|$ is $\leq 2\cdot\text{avg}(|C_i|)$.

{\paragraph{Random Permutation}
To further constrain the size of the aggregated prediction sets, we utilize an observation regarding random permutations \cite{gasparin2024}. We demonstrate our approach in Algorithm \ref{algo:randomPermutation}. }We aggregate prediction sets $C_1, \dots, C_n$ by randomly permuting their indices and construct the final prediction set $C^\pi$
based on majority voting criteria similar as in Theorem 1. Specifically, we iteratively count occurrences of labels 0 and 1 across each element in the permutation, considering all prefixes of the permutation ending in index $i$ (for $1 \leq i \leq n$) adding an element to the prediction for this prefix $C^\pi(\pi_1 : \pi_i)$ if its occurrence meets or exceeds half of the current iteration index $i$. The final result $C^\pi$ contains elements that are supported by each prediction set across all prefixes.

\begin{algorithm}
\caption{Random Permutation Method}
\begin{small}
\begin{algorithmic}[1]
\Require Prediction Sets  $C_1, \dots, C_n$
\Ensure An aggregated prediction set $C^\pi$
\State $C^\pi \gets \emptyset$
\State $\pi_1 \dots \pi_k \gets$ random permutation of $\{1, 2, \ldots, n\}$;
\For {each $i$ in range $1\dots k$}
\State $C^\pi(\pi_1 : \pi_i) \gets \emptyset$
\State $Count_0 \gets $ count number of occurrences of 0 in $C_{\pi_1} \dots C_{\pi_i}$
\State $Count_1 \gets $ count number of occurrences of 1 in $C_{\pi_1} \dots C_{\pi_i}$
\If{$Count_0\geq \frac{i}{2}$}
\State $C^\pi(\pi_1 : \pi_i) \gets C^\pi(\pi_1 : \pi_i) + \{0\}$ 
\EndIf
\If{$Count_1\geq \frac{i}{2}$}
\State $C^\pi(\pi_1 : \pi_i) \gets C^\pi(\pi_1 : \pi_i) + \{1\}$
\EndIf
\State $C^\pi \gets C^\pi \cap C^\pi(\pi_1 : \pi_i)$
\EndFor
\end{algorithmic}
\end{small}
\label{algo:randomPermutation}
\end{algorithm}


We show that the worst-case marginal probability guarantee for Algorithm \ref{algo:randomPermutation} is the same as that of Theorem \ref{th:1} for the case of $\theta=1/2$; however, the size of the prediction set it produces could be smaller.

\begin{theorem}
 Let $C_1, \ldots, C_n$ be prediction sets according to Equation \ref{eq:conf}. Then,
    $$p(c^* \in C^\pi) \geq 1-2\alpha$$ and $|C^\pi| \leq |C^\theta|$ when $\theta = \frac{1}{2}$.
\end{theorem}
\begin{small}
\begin{proof}
Let $\phi_k = \mathbf{1} \{ c \in C_k \}$. By conformal prediction, we know $E[\phi_k] \leq \alpha$.
\begin{align*}
p(c^* \notin C^\pi) = p(\exists k \leq K: c \notin C^\pi(\pi_1 : \pi_k))\\
= p(\exists k \leq K: \frac{1}{k} \sum_{i=1}^{k} \phi_{\pi_i} \geq \frac{1}{2}) \leq 2E[\phi_i] \leq 2\alpha
\end{align*}
where the first inequality follows the exchangeable Markov inequality (EMI)\cite{ramdas2023}. The second part $|C^\pi| \leq |C^\theta|$ is obvious as $C^\pi(\pi_1 : \pi_n)$ is the same as $C^\theta$. 
\end{proof}
\end{small}

Since Algorithm \ref{algo:randomPermutation} provides the same probability guarantees as in Theorem \ref{th:1} with a smaller prediction set compared to majority voting, we employ it to aggregate the results over mBPP, and  we label the token as a branching point if and only if $1$ appears in the final prediction set $C^\pi$.

\subsection{Abstention Mitigation}
\label{sec:mit}
Upon identification of a branching point, the model can be explicitly directed to cease further generation. This approach ensures the enforcement of the analytical results presented in Section \ref{sec:bppsec}. Subsequently, we explore methodologies to mitigate abstentions, either through process self-correction or by soliciting human intervention.
Algorithm \ref{algo:traceBack} is initiated following the identification of a branching point and yields a list of tables for inspection. The algorithm accepts three inputs along with the schema linking model: the set of potential tables $\mathcal{T}$ for linking, the sequence of tokens generated up to that point, and the identified branching point. It utilizes a {\bf decode} function that takes as arguments a sequence $S$ of tokens and the set of tables $\mathcal{T}$. The decode function operates by concatenating the tokens in $S$ from left to right to form table names, which are then compared against the tables in $\mathcal{T}$. The list of tables returned is determined by  the set difference of the  {\bf decode} result computed on the sequence before and after the branching point. When applying {\bf decode} in the sequence after the branching point, if a token subsequence of $S$ cannot be matched to any table name (note that such a subsequence will invariably be a suffix of $S$), we request that the model continues generation until a next table in $\mathcal{T}$ is identified by {\bf decode}.




\begin{algorithm}
\caption{Table Trace Back}
\begin{small}
\begin{algorithmic}[1]
\Require A branching point token $x_b$, Set of generated tokens $X = x_1 \dots x_b$, Set of possible tables $\mathcal{T}=\{T_1 \dots T_{\nu}\}$, Schema Linking Model $M$
\Ensure $\mathcal{T}_b \subset \{T_1 \dots T_{\nu}\}$ The corresponding possible erroneous tables.
\State $\mathcal{T}_{pre} \gets$ {\bf decode} $X[:-1] = x_1 \dots x_{b-1}$ into a set of tables $\mathcal{T}_{pre} \subset \mathcal{T}$
\State $\mathcal{T}_{after} \gets$ {\bf decode} $X$ into a set of tables $\mathcal{T}_{after} \subset \mathcal{T}$
\While {$\mathcal{T}_{after} - \mathcal{T}_{pre} = \emptyset$}
\State $x_{new} \gets $ generate the next token using $M$ and $X$
\If{$x_{new} = $ eos}
\State \Return $\mathcal{T}_b \gets \mathcal{T}[-1:]$ 
\EndIf
\State $X \gets X + x_{new}$
\State $\mathcal{T}_{after} \gets$ {\bf decode} $X$ into a set of tables $\mathcal{T}_{after} \subset \mathcal{T}$
\EndWhile
\State \Return  $\mathcal{T}_b \gets \mathcal{T}_{after} - \mathcal{T}_{pre}$
\end{algorithmic}
\end{small}
\label{algo:traceBack}
\end{algorithm}

\paragraph{{\bf Surrogate Filter}}
Upon identification of a branching point, a potential corrective action involves attempting to rectify the erroneous token prediction. Algorithm \ref{algo:traceBack} yields a set of one or more tables to which the branching point is attributed. These tables may either indicate a source of ambiguity within the schema or suggest that the model erroneously predicts these tables, consequently impeding further generation.

To automatically continue the 
{schema linking} process, we propose the following mitigation strategy:
Let $\mathcal{T}_b$ 
denote the set of tables returned by Algorithm \ref{algo:traceBack}. We utilize a surrogate model\footnote{Though ``surrogate model" typically conveys a distinct meaning within the domain of machine learning, here it denotes a stand-in for a human expert.}  (Deepseek-7B) to classify whether the tables in $\mathcal{T}_b$ are indeed relevant/irrelevant to the query under consideration. We fine-tune the surrogate model utilizing the BIRD and Spider training datasets, specifically for the purpose of providing answers to the following classification problem: Given a schema and a query, is a provided set of tables relevant to the query or not? We formulate the following prompt using $\mathcal{T}_b$, the {\tt Schema}, and the {\tt Question} as input to the surrogate model.
\begin{small}
\begin{lstlisting}
{Schema}
Question: {Question}
Is the  `{$\mathcal{T}_b$}` relevant to the question: 
(A) True (B) False
\end{lstlisting}
\end{small}

The {schema linking} process halts only if the surrogate model explicitly confirms the irrelevance of these tables by outputting ``False''; otherwise it continues.
Notice that the same process can be utilized for columns as well.

\paragraph{{\bf Soliciting Human Feedback}}

Instead of using the surrogate model, upon identification of branching points, if correction is feasible, we could enable the continuation of token generation, potentially leading to the correct answer. This observation motivates the development of an interactive system wherein the model performs self-corrections based on human feedback.
The viability of this framework relies on the assumption that the human participant will consistently provide valuable assistance by offering accurate information in response to queries posed by the schema linking model.

When a token is identified as a branching point, we employ Algorithm \ref{algo:traceBack} to trace back to the corresponding tables, yielding a set $\mathcal{T}_b$. Subsequently, we initiate user interaction.
For each table $T_i\in \mathcal{T}_b$,  we solicit user confirmation regarding its relevance to the query. If the user affirms the table's relevance, 
token generation proceeds without further intervention, utilizing the tokens in $T_i$. Conversely, if the user deems all tables $T_i \in \mathcal{T}_b$ irrelevant, we request the correct table name. 
The provided table name and its associated tokens are then utilized to continue the generation process.

\section{Experimental Evaluation} \label{sect:exp}

In this section, we present the results of a detailed experimental evaluation assessing the impact of \OurMethod ~both in schema linking accuracy and subsequent SQL generation. We detail our experimental environment, benchmark datasets utilized, evaluation metrics, and methodology. We stress that \OurMethod ~is a framework that can be utilized by {\em any} schema linking methodology utilizing a transparent LLM. For that reason, we keep the schema linking model simple and focus on the evaluation of the impact of \OurMethod.
{
In this work, we assume that the training distribution aligns with the testing distribution. Addressing performance consistency under data drift, which may require alternative approaches like transfer learning, is outside the scope of this paper.
}

\subsection{Experimental Methodology}

We perform empirical evaluations using two relational schema-to-SQL query generation benchmarks: Spider \cite{yu2018spider} and BIRD \cite{li2024can}. Adhering to established experimental protocols from prior literature, we utilize the training corpus to train our model and assess its efficacy on the validation and test datasets. However, our research objective centers on enhancing the robustness of natural language to SQL query translation. Given that our methodology produces output with a non-standard schema (incorporating abstention), incompatible with the test set submission specifications, we are unable to evaluate our approach on the held-out test data. As a result, our primary performance metrics and reliability assessments are conducted on the publicly accessible validation set { and test set for Spider 1.0}, which allows for more granular analysis. { 
Moreover, training each BPP requires a labeled dataset, which is readily available in benchmarks. In operational database settings, one can enhance database logs—containing at least the SQL query and, potentially, the result—by adding query descriptions through human curation or by generating them directly using an LLM.}

All the experiments were conducted on a high-performance workstation. The workstation is equipped with a 16-core AMD Ryzen 5955W CPU and a single NVIDIA A100 GPU with 80 GB of GPU RAM.

\subsubsection{Benchmark Datasets}
We utilize two popular text-to-SQL benchmarks for our evaluation.

\paragraph{\textbf{Spider}}
The Spider benchmark is a comprehensive dataset for evaluating text-to-SQL generation models, containing a training set with 8,659 samples, a development set with 1,034 samples, and a hidden test set. It includes 200 databases with different schemas and domains, along with complex SQL queries paired with natural language questions. The benchmark challenges models to generalize across diverse databases and handle intricate SQL structures,  and it is widely used in research to advance natural language interfaces to databases.

\paragraph{\textbf{BIRD}} 
The BIRD benchmark evaluates text-to-SQL systems with a focus on large-scale and more varied data. It includes a training set of 9,428 samples, a development set with 1,534 samples, and a hidden test set. BIRD encompasses 95 databases across 37 distinct professional domains and presents a substantial challenge with ``dirty'' values that retain the original, often non-standard format from real-world scenarios. Moreover, BIRD offers external knowledge for specific samples to facilitate the generation of the correct SQL query. Consequently, text-to-SQL parsers must not only analyze these non-standard formats before engaging in reasoning but also accurately incorporate external knowledge into the text-to-SQL generation process.

\subsubsection{Baseline Models}
Due to the unique setting of \OurMethod ~(namely introducing abstention in the text to SQL process) we opt for a simple transparent LLM-based schema linking model. We stress that any such model can be readily utilized with our approach. Thus we focus on how effective and accurate the abstention mechanisms we introduce are and their impact on the ensuing schema linking task. { Our primary objective is to enhance the reliability of any schema-linking model by incorporating abstention mechanisms, while maintaining the same level of accuracy as traditional models that do not utilize abstention. Given our hardware constraints, we currently employ simple, fine-tuned schema-linking models at a scale feasible to us. However, we emphasize that the choice of the base schema linking model is orthogonal to our approach; larger-scale models, such as the 236B DeepSeek-Coder, will also benefit from the RTS framework, which not only improves overall accuracy but also significantly enhances reliability by allowing models to abstain when uncertain. }

However, we emphasize that the choice of the base schema-linking model is independent of our approach, and we assert that larger models, such as the 236B DeepSeek-Coder, would also benefit from the RTS framework. This framework not only enhances overall accuracy but also greatly improves reliability by enabling models to abstain when uncertain

For this reason, we utilize a simple finetuned schema linking model, stressing that the choice of the base schema linking model is orthogonal to our approach. 

\paragraph{\textbf{Schema Linking Models}} Our approach employs large language models specifically for table linking and column linking. We use Deepseek-7B \cite{deepseek-llm} as the schema linking model since it demonstrates the best performance among small-scale models \cite{pourreza2024dts}. We employ supervised fine-tuning of Deepseek-7B on the schema linking task using the corresponding training data for each benchmark and evaluate its performance on the development dataset. Several papers \cite{lei2020re, pourreza2024din} present ideas on how to improve schema linking accuracy on top of fine tuning (e.g., retrieval of sample column values) and these can be readily adapted here as well without affecting the operation of \OurMethod.


\paragraph{\textbf{Text-to-SQL Model}} The choice of text-to-SQL model is also orthogonal to \OurMethod, as we focus on generating more reliable schema linking. To demonstrate the effect of our approach, we consider two interpretable LLMs: Deepseek-7B \cite{deepseek-llm} and CodeS \cite{li2024codes}. CodeS is an open-source LLM specifically tailored for text-to-SQL generation, achieving superior accuracy with a much smaller parameter size. For both Deepseek-7B and CodeS, we conducted supervised fine-tuning on their respective training datasets, ensuring that the models were well-adapted to the text-to-SQL generation for each benchmark.

\paragraph{\textbf{Implementation Details}}
Unless stated otherwise, we set the error level $\alpha$ to 0.1. Given that the LLM implementing the schema linking
model has $n$ layers, we select the $k$ best performing sBPP classifiers to form the mBPP. To assess the quality of a sBPP we compute the AUC scores over the calibration dataset. Unless stated otherwise, we use $k=5$ in our experiments. {Additionally, we train each BPP using approximately 10\% of the training set, which has consistently yielded good performance.}
For all LLMs, we use greedy decoding with a temperature setting of 0 to ensure reproducibility.

\subsection{Evaluation Metrics}
\paragraph{\textbf{Schema Linking Model}}

In accordance with previous work \cite{pourreza2024dts}, the schema linking model is evaluated based on exact set match, precision, and recall. Taking table linking as an example, let $\mathcal{T}_i $ be the set of ground truth tables for a specific query and $\hat{\mathcal{T}}_i $ be the set of predicted tables. Exact Set Match (EM), measures the percentage of instances where the predicted set exactly matches the ground truth, $\frac{1}{n}\sum_{i=1}^n\mathbf{1}(\mathcal{T}_i =\hat{\mathcal{T}}_i)$. Precision is calculated as the number of correct tables predicted (true positives) divided by the total number of tables predicted (true positives plus false positives), $\frac{1}{n}\sum_{i=1}^n\frac{|\mathcal{T}_i \cap\hat{\mathcal{T}}_i|}{|\hat{\mathcal{T}}_i|}$. Recall is calculated as the number of correct tables predicted (true positives) divided by the total number of tables in the ground truth (true positives plus false negatives), 
$\frac{1}{n}\sum_{i=1}^n\frac{|\mathcal{T}_i \cap\hat{\mathcal{T}}_i|}{|\mathcal{T}_i|} $.


\begin{table}[h]
\centering
\caption{Schema Linking Model Performance}
\begin{tabular}{lccccc}
\toprule
Type  & Dataset& Exact Match (\%)   & Precision (\%)  & Recall (\%)  \\
\midrule
Table & Bird  &  79.70 & 92.85 & 95.00 \\
Column & Bird &  75.32 & 89.87 & 88.79 \\
Table & Spider-dev  &  93.71 & 98.17 & 96.95 \\
Column & Spider-dev &  88.98 & 94.41 & 94.09 \\
Table & Spider-test  &  92.72 & 97.64 & 96.74 \\
Column & Spider-test &  87.99 &  92.21 & 93.02 \\
\bottomrule
\end{tabular}
\label{tab:LLMSchemaLinker}
\vspace{-1em}
\end{table}

\paragraph{\textbf{Branching Points}}
We utilize the AUC score, which measures the ability of the model to distinguish between the positive and negative classes, to evaluate the performance of each Branching Point Predictor (sBPP).

To evaluate the performance of the conformal prediction with multiple BPPs, (mBPP) we consider the following metrics: coverage and extra abstention rate (EAR). Let $S = \{s_1 \dots s_n\}$ denote the ground truth labels for each token and $\{\hat{s}_1 \dots \hat{s}_n\}$ to be the predicted labels. The coverage and EAR are defined as follows
\begin{enumerate}
    \item Coverage is calculated as the percentage of branching points correctly detected by mBPP among all the branching points
    $$Coverage = \frac{|\{i| \hat{s}_i = s_i = 1, i \in 1,2\dots n \}|}{|\{i| s_i = 1, i \in 1,2\dots n \}|}$$
    \item Extra abstention rate is calculated as the percentage of incorrect predictions of branching points among the entire dataset. It is an indication of the percentage of unnecessary abstention. 
    $$EAR = \frac{|\{i| \hat{s}_i = 1, s_i = 0, i \in 1,2\dots n \}|}{n}$$
\end{enumerate}
Ideally, we aim to achieve higher coverage with only a small EAR. Increasing $\alpha$ could increase coverage; however, it might simultaneously increase EAR as non-branching points are more likely to be identified as branching points. Therefore, we primarily focus on studying the trade-off between coverage and EAR for BPP.

\paragraph{\textbf{Schema Linking With Abstention}}
To evaluate \OurMethod, we consider three key metrics: exact set match, false abstention rate, and true abstention rate. For the case of table linking as an example (predicting tables relevant to a query q), let $\hat{s}_1 \dots \hat{s}_n$ denote the decision of whether to abstain on the test sample. Let $T_1 \dots T_q$ denote the ground truth tables for a query q and $\hat{T}_i \dots \hat{T}_q$ be the predicted tables. Then the true abstention rate (TAR) is calculated as the percentage of instances where the model abstains from making a prediction and is not capable of making the correct one (namely if we allow the model to predict it will make the wrong prediction). $$TAR = \frac{|\{i|\hat{s}_i = 1,T_i = \hat{T}_i \}|}{n}$$

In contrast, the false abstention rate (FAR) is calculated as the percentage of instances where the model abstains from making a prediction despite being capable of making a correct one. 
$$FAR = \frac{|\{i|\hat{s}_i = 1,T_i \neq \hat{T}_i \}|}{n}$$

In other words, TAR measures the percentage of abstentions that correctly capture instances where the model cannot make a correct prediction, while FAR measures the unnecessary abstentions incurred due to detection errors. The exact set match represents the accuracy of predictions in which the model does not abstain..




\begin{table}[h]
    \centering
    \begin{minipage}{0.45\textwidth}
        \centering 
        \caption{Average sBPP AUC (\%) for Bird and Spider Dataset}
        \begin{small}
        \begin{tabular}{lccc}
        \toprule
        Type & Bird & Spider-dev & Spider-test\\
        \midrule
        Table & 97.16  & 98.43& 97.90 \\
        Column & 96.70 & 96.90& 96.60 \\
        \bottomrule
        \end{tabular}
        \end{small}
        \label{tab:SBPP}
    \end{minipage}%
    \hfill
    \begin{minipage}{0.45\textwidth}
        \centering
        \caption{Surrogate Model Accuracy (\%) for Bird and Spider Dataset}
        \begin{small}
        \begin{tabular}{lccc}
        \toprule
        Type & Bird & Spider-dev &  Spider-test\\
        \midrule
        Table & 92.37  & 96.45 & 96.02\\
        Column & 94.06 & 96.30 &  96.00\\
        \bottomrule
        \end{tabular}
        \end{small}
        \label{tab:Surrogate}
    \end{minipage}
    \vspace{-1em}
\end{table}

\paragraph{\textbf{Evaluating Text-to-SQL}}

We evaluate the downstream text-to-SQL model using execution accuracy (EX) \cite{yu2018spider, li2024can}. Execution accuracy assesses the correctness of SQL queries generated by the model, comparing their results with ground truth results produced from golden query execution on the database, which has been widely adopted by many previous works and benchmarks as the standard metric.

\subsection{Experimental Results}
\paragraph{\textbf{Evaluating Individual Components}}
In this section, we demonstrate the effectiveness of \OurMethod ~through experiments. We start with evaluating the key components of \OurMethod, namely the schema linking model, sBPP, and the surrogate model.

{The results\footnote{ We include the results from the Spider development and test sets in the experiment. As its performance remains consistent throughout, we refer to them as the Spider dataset in the following analysis.} in Table \ref{tab:LLMSchemaLinker} display the baseline performance of the schema linking model we adopt across two datasets: BIRD and Spider.} The schema linking model performs better for both tables and columns on the Spider dataset compared to the more challenging dataset BIRD.  Admittedly, the schema linking model we adopt may not provide the best schema linking accuracy compared to other available approaches \cite{li2023resdsql,talaei2024} that utilize additional metadata (e.g., matching sample data, checking datalog) or involve retraining a new language model. However, we emphasize that our goal is not to maximize the basic accuracy of schema linking. Instead, we introduce a mechanism that allows the model to abstain from predictions in certain cases, thereby enhancing overall robustness and adaptability. Our proposal can be adopted by any transparent schema linking model. Therefore, the choice of schema linking model is orthogonal to our proposal.

To evaluate the effectiveness of sBPP, we calculate the average AUC (Area Under the Curve) score for all the sBPP models used in conformal prediction. Table \ref{tab:SBPP} summarizes the results for BIRD and Spider datasets. As is evident from the table, sBPP achieves a near-perfect AUC score for both datasets, suggesting that the sBPP models are effective in this context. 

Lastly, we use a surrogate model as a binary classifier (to assess whether tables/columns are relevant to a query given a schema description) and evaluate its classification accuracy on the development set of the BIRD and Spider benchmarks in Table \ref{tab:Surrogate}. As presented in the table, the accuracy is high (>94\%) for both benchmarks. However, it should be noted that the accuracy presented is for classifying the relevance of tables or columns, for the purposes of continuing the generation, which does not correspond to the schema matching accuracy using this model. 

\begin{table*}[h]
\centering
\caption{\OurMethod ~Schema Linking Model Performance}
\begin{small}
\resizebox{\textwidth}{!}{%
\begin{tabular}{lcccccccccc}
\toprule
Method & Type & \multicolumn{3}{c}{Bird} & \multicolumn{3}{c}{Spider-dev} &\multicolumn{3}{c}{Spider-test} \\
       &      & EM (\%)& TAR (\%)& FAR (\%)& EM (\%)& TAR (\%)& FAR (\%)& EM (\%)& TAR (\%)& FAR (\%)\\
\midrule
\multirow{2}{*}{mBPP-Abstention} & Table & 98.89 & 19.10 & 12.77 & 99.86     & 6.51   &  5.27 & 99.67     & 6.28   &  4.98\\
& Column& 97.38 & 22.01 & 13.53 &97.73 & 8.75   & 7.46 &97.52 & 9.25   & 8.32\\
\midrule
\multirow{2}{*}{Surrogate filter} & Table  & 90.80 & 10.90 & 2.2 &  96.77 &  3.05  & 1.70 &  95.47 &  4.10  & 2.03\\
 & Column & 89.76 & 14.34  & 5.98 & 92.71  &  3.70  & 3.35 &90.18  &  4.63  & 4.12\\
\midrule
\end{tabular}%
}
\label{tab:Abstention Schema Linker}
\end{small}
\vspace{-1em}
\end{table*}

\begin{table}[h]
\centering
\caption{Schema Linking Performance with Human Feedback}
\begin{small}
\begin{tabular}{lcccc}
\toprule
Dataset& Type   &EM (\%) & TAR (\%)&FAR (\%)\\
\midrule
Bird & Table  & 96.90  & \multirow{2}{*}{18.95}   &\multirow{2}{*}{13.65}   \\
Bird & Column & 96.02  &        &   \\

Spider-dev & Table  & 98.93   & \multirow{2}{*}{6.46} &\multirow{2}{*}{8.15}     \\
Spider-dev & Column & 96.71   & &  \\
Spider-test & Table  & 99.02   & \multirow{2}{*}{6.61} &\multirow{2}{*}{8.20}     \\
Spider-test & Column & 96.11   & &  \\
\bottomrule
\end{tabular}
\label{tab:Correction Schema Linker}
\end{small}
\vspace{-1em}
\end{table}

\paragraph{\textbf{mBPP with Abstention and Surrogate Model}}
We first consider the case where the model abstains once a branching point is detected by mBPP. In this scenario, we test the schema linking model for table linking and column linking independently (i.e. column linking will not receive input from the table linking result). This is to minimize the dependency between these two processes and evaluate them independently. Later, we will demonstrate their joint effect (i.e. identifying the relevant table first, then finding the column for each table) and the implications on the abstention rate. Table \ref{tab:Abstention Schema Linker} summarizes the results. 

Looking at the first row (mBPP-Abstention) of table \ref{tab:Abstention Schema Linker}, both table-level and column-level results demonstrate high exact match score for all the cases where the model decides not to abstain for both BIRD and Spider (achieving EM of 98.89\% and 97.38\% respectively). However, this method also has high FAR, which indicates that roughly more than 12.77\% of tables and 13.53\% columns during inferences exhibit unnecessary abstentions for BIRD. For Spider the numbers are 5.27\% for tables and 7.46\% for columns. To put into perspective out of 100 queries posed by BIRD, we will decide to conduct table linking on 68.13\% of those achieving an accuracy of 98.89\% in table linking, and abstain on the rest of the queries. Similarly, in 64.46\% of those queries we will conduct column linking achieving 97.38\% exact match score and abstain on the rest.

Looking at the second row of Table \ref{tab:Abstention Schema Linker} we analyze the impact of utilizing the surrogate model. In this case, FAR is reduced significantly, but both EM and TAR are reduced as well. 
It appears that the application of the surrogate model is able to identify correctly cases when the model should not abstain (thus reducing FAR). However at the same time it appears that the surrogate model assess (wrongly) that we should not abstain in some instances thus forcing the schema linking model to generate incorrect output (reducing EM and TAR).


In summary, in practice both methods (mBPP-Abstention and the application of surrogate model) achieve better EM, compared to the theoretical error-level $\alpha$ (80\% in this case using $\alpha=0.1$). The surrogate model introduces a trade-off between EM and FAR. While it reduces unnecessary abstentions, it also limits the overall EM by failing to recognize instances beyond the surrogate model's capabilities. 

\paragraph{\textbf{mBPP with Human Feedback}}
We now discuss \OurMethod ~assuming the ability to solicit human assistance upon encountering branching points. In this approach, we consider a schema linking model that conducts both table and column predictions by first identifying the relevant table, followed by identifying the appropriate columns for each table. In this case, we abstain if either the table linking or column linking chooses to abstain. Note that if the table and column linking abstain on entirely different instances, the FAR and TAR obtained in this joint process should be proportional to the sum of the FAR and TAR of each component.

Table \ref{tab:Correction Schema Linker} presents the results when we solicit human feedback each time a branching point is detected. Since table and column linking takes place jointly we compute TAR and FAR metrics jointly and for this reason, there is one value for TAR and FAR for each benchmark dataset. We achieve at least 96\% accuracy for table and schema linking in both benchmarks.
Notice that the FAR value is also low, pointing to a small number of cases in which a human is involved unnecessarily (namely the model can make the correct prediction independently). Comparing the results for Spider with those on BIRD, the numbers for Spider are slightly better as Spider is considered an easier benchmark.

Comparing the results of table \ref{tab:Correction Schema Linker}, with those in Table \ref{tab:Abstention Schema Linker} relevant to mBPP-Abstention (first row) we observe that the False Abstention Rate (FAR) and True Abstention Rate (TAR) for both BIRD and Spider are much lower than the sum of the FAR and TAR obtained when table and column linking are considered independently. 
This indicates a significant overlap in the inferences where either linking operation abstains. Specifically, if the table linking operation abstains, the column linking operation is likely to do the same. This is expected because if the schema linking model, that links only tables is uncertain about a table's relevance to the question, the schema linking model linking only columns will also struggle to link the corresponding column.


\begin{table}[]
\centering
\caption{Execution Accuracy (\%) for Downstream Text-to-SQL Model with Different Schemas}
\begin{small}
\begin{tabular}{lccccc}
\toprule
Model & Schema Type & Bird & Spider-dev & Spider-test\\
\midrule
\multirow{3}{*}{Deepseek-7B} &  Golden Schema & 66.21 & 90.13 & 90.02 \\
                              & \OurMethod-Schema & 64.72 & 88.90 & 88.20 \\
                              & DTS-SQL\cite{pourreza2024dts} & 55.8 &  85.50 &84.4\\
\midrule
\multirow{4}{*}{CodeS-15B} & Golden Schema & 66.27 & 90.02 & 90.10\\
                            & \OurMethod-Schema & 65.19 & 89.10 & 88.68 \\
                            & CodeS\cite{li2024codes} & 58.47 & 84.90 & 85.01\\
\bottomrule
\end{tabular}
\end{small}
\label{tab:text-to-SQL}
\end{table}

\paragraph{\textbf{Text-to-SQL}}
Finally, we demonstrate the performance of text-to-SQL with the schema derived from \OurMethod, which solicits human feedback in downstream LLMs. We also compare the results of utilizing a perfect schema linking model, where the schema contains only the relevant tables and columns provided to the fine-tuned SQL generators. Additionally, we compare these results with the best-reported methods utilizing this model (i.e., DTS-SQL for DeepSeek-7B and CodeS for CodeS-15B). The results are presented in Table~\ref{tab:text-to-SQL}. We note that the results utilizing the golden schema present an upper bound for this text-to-SQL model. It is evident that this upper bound score is much higher than the corresponding best-reported method,  highlighting the importance of correct schema linking. With human assistance, we are able to achieve near-perfect schema linking (as reported in Table \ref{tab:Correction Schema Linker}) and hence attain an EX value comparable to that of the upper bound. We stress that the EX value we achieve with \OurMethod-Schema and DeepSeek-7B is 65.19\% compared to 66.95\% of the current SOTA technique at the BIRD leader board utilizing Gemini, a model orders of magnitude larger than Deepseek-7B. It is important to highlight that during this process, the human we solicit assistance from does not need to write SQL queries; it is desirable to be familiar with the schema to assess tables and columns relevant to a query when asked.

\begin{figure}[]
    \centering
    \begin{subfigure}[b]{0.23\textwidth}
        \centering
        \includegraphics[width=\textwidth]{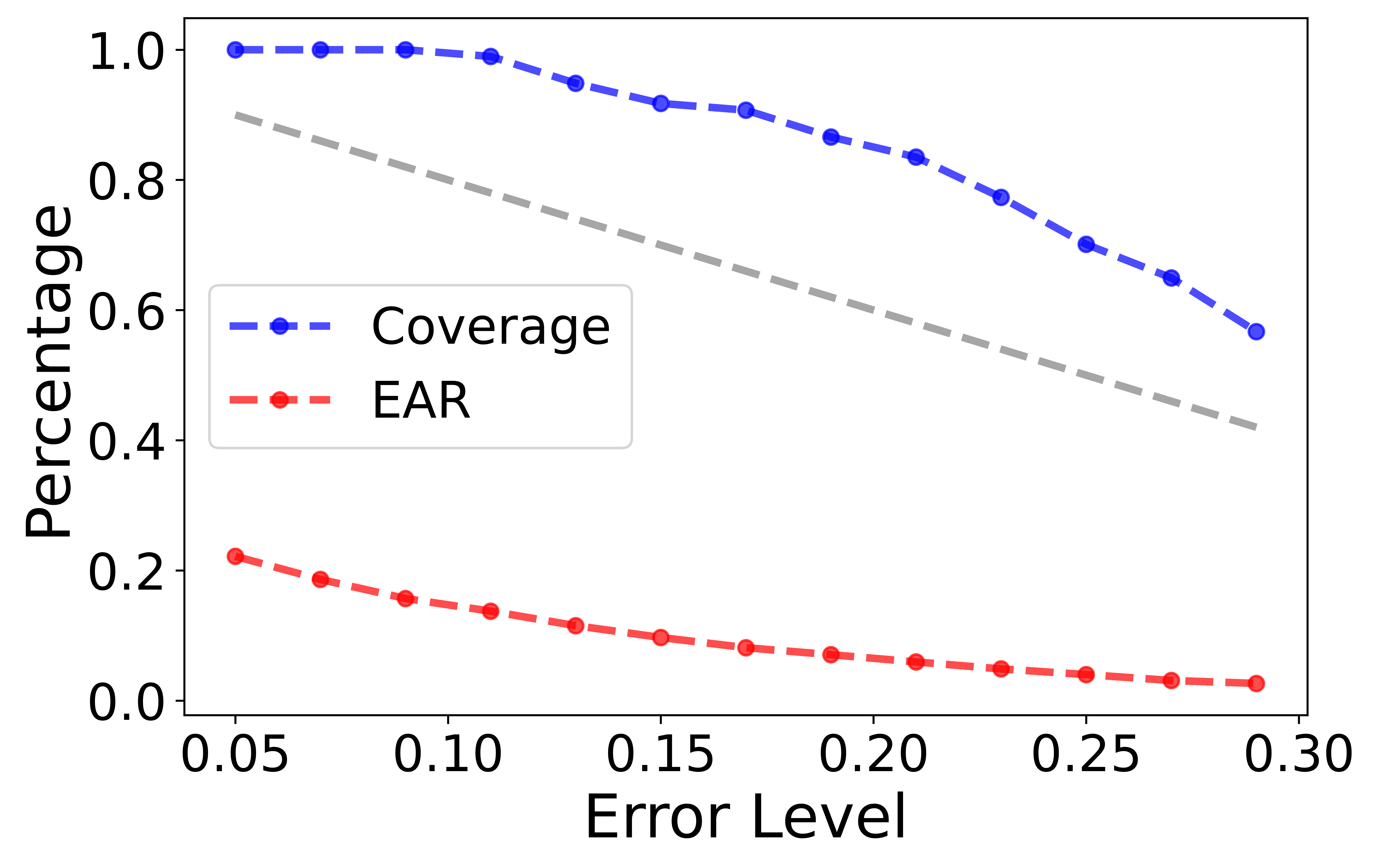}
        \caption{Table Linking mBPP}
    \end{subfigure}
    \hfill
    \begin{subfigure}[b]{0.23\textwidth}
        \centering
        \includegraphics[width=\textwidth]{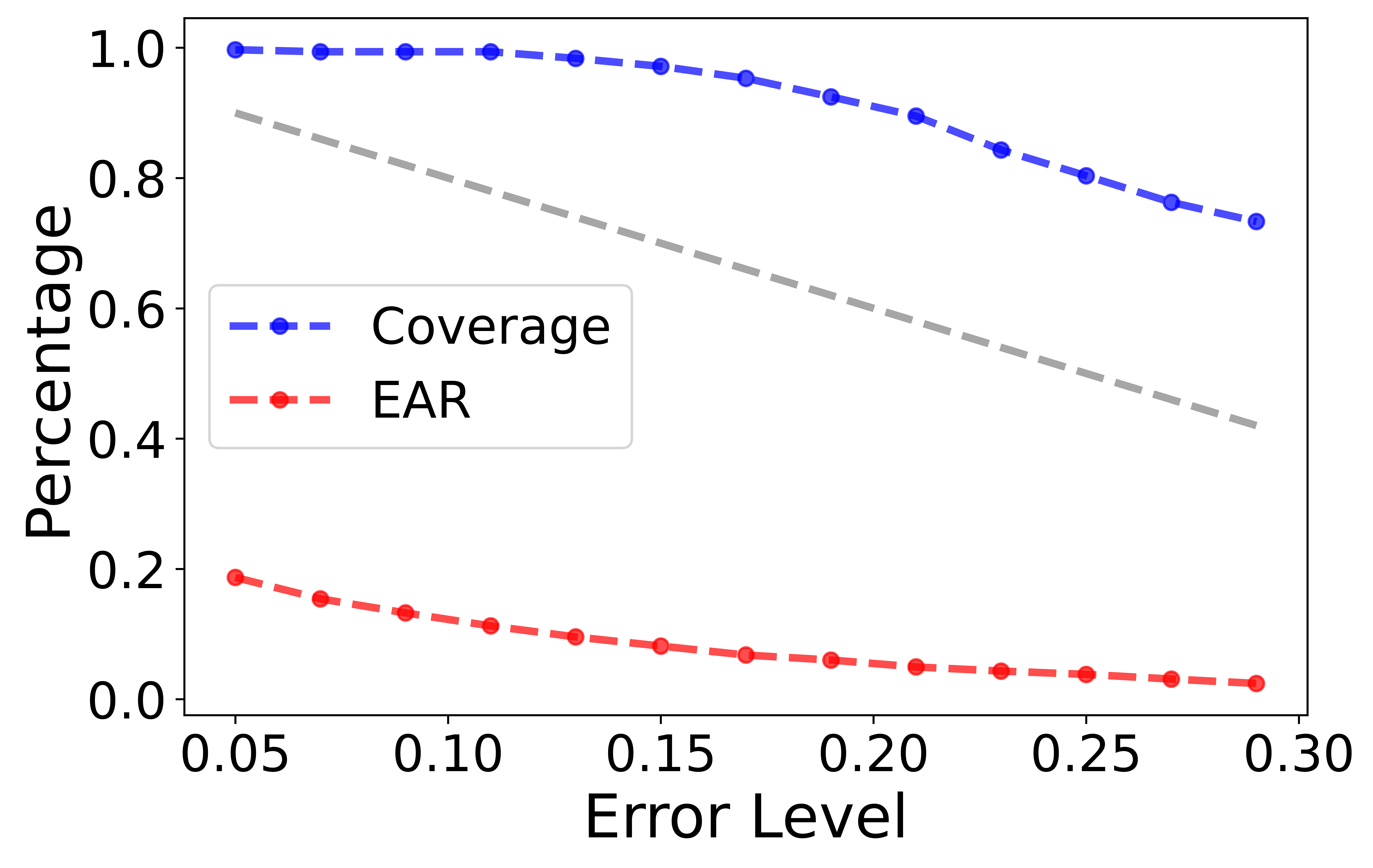}
        \caption{Column Linking mBPP}
    \end{subfigure}
    \vspace{-1em}
     \caption{Coverage v.s EAR for each error level. Grey line indicates the theoretically guaranteed error level.}
    \label{fig:coverage}
\end{figure}

\paragraph{\textbf{Ablation Study}}
We conduct an ablation study, utilizing the BIRD dataset for brevity; we stress that similar trends are observed in the Spider dataset.
\paragraph{The effect of $\alpha$}
We present the coverage and EAR varying $\alpha$ in Figure \ref{fig:coverage}. As shown in the figure, the empirical coverage of mBPP consistently exceeds the theoretical threshold $\alpha$. The theoretical guarantee is presented as a dotted black like in the figure always enveloped by coverage. Moreover, the empirical coverage remains relatively constant when the threshold is small (< 0.15). This behavior indicates that mBPP provides reliable error quantification, especially for lower threshold values.

\begin{figure}[]
    \centering
    \begin{subfigure}[b]{0.23\textwidth}
        \centering
        \includegraphics[width=\textwidth]{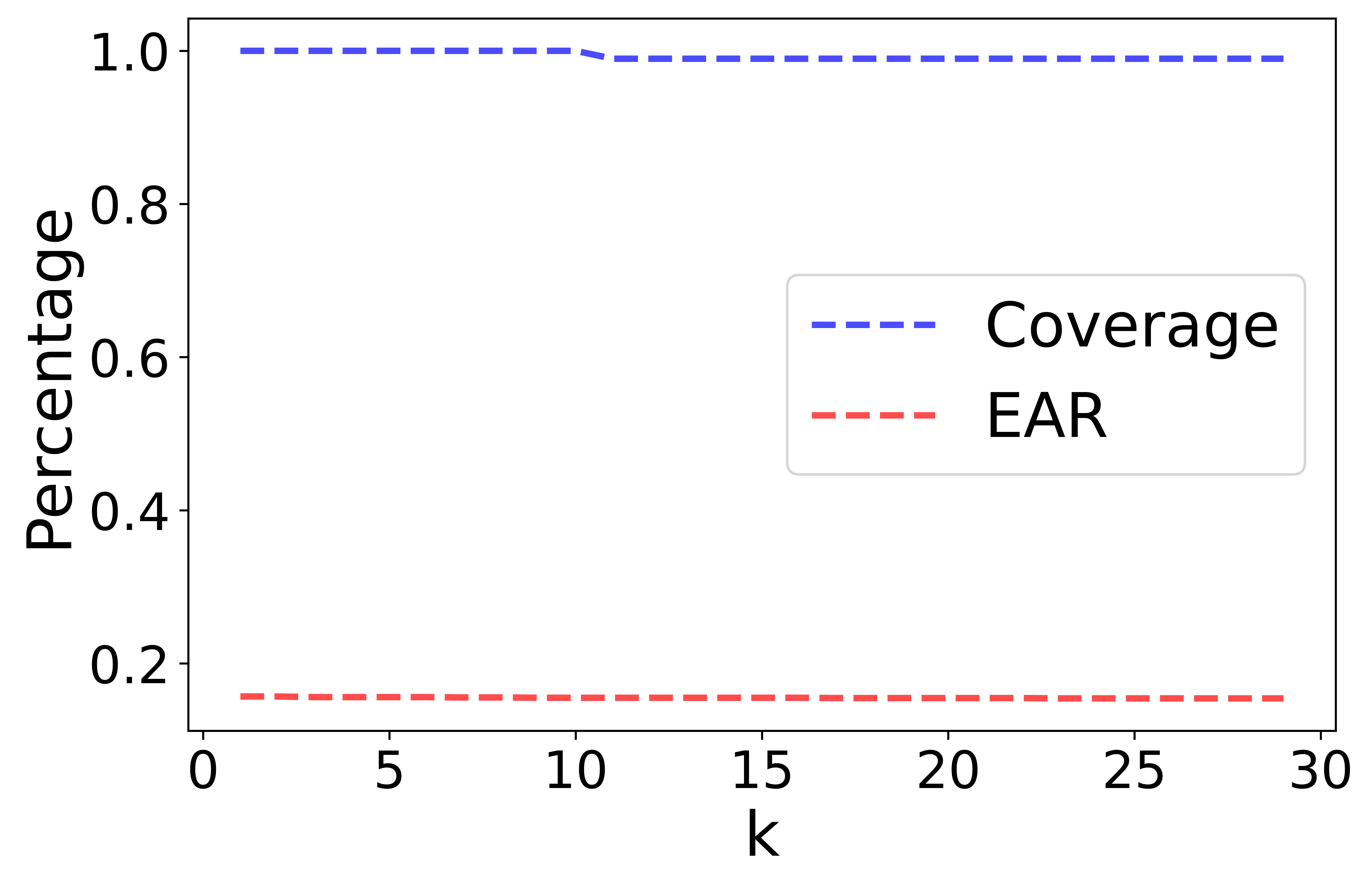}
        \caption{Random Permutation}
    \end{subfigure}
    \hfill
    \begin{subfigure}[b]{0.23\textwidth}
        \centering
        \includegraphics[width=\textwidth]{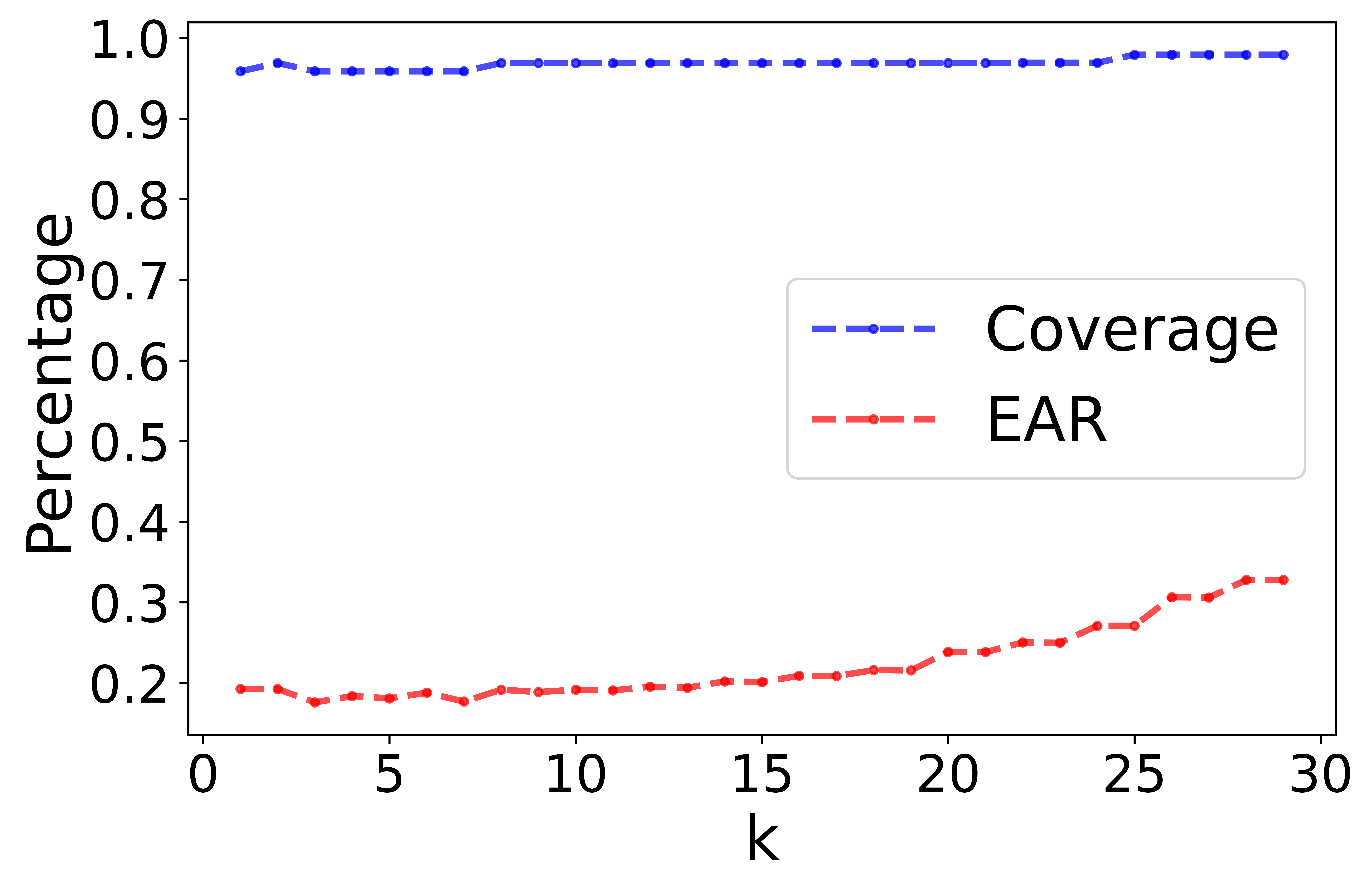}
        \caption{Majority Vote}
    \end{subfigure}
    \vspace{-1em}
     \caption{Coverage vs. EAR for different $k$ values}
    \vspace{-1em}
    \label{fig:k}
\end{figure}

\paragraph{The effect of $k$ and random permutation}
Lastly, we study the effect of $k$ (number of sBPP utilized in mBBP) along with the effects of random permutation utilized in Algorithm \ref{algo:randomPermutation}. In this ablation study, we fix the error level at $\alpha=0.1$ and use table linking for demonstration. A similar trend can also be observed for column linking. In Figure \ref{fig:k}, we plot the coverage and EAR curves for different values of $k$. We consider two methods, majority vote (as per Theorem \ref{th:1}) and random permutation (Algorithm \ref{algo:randomPermutation}), to aggregate the results of each sBPP. As demonstrated in the figure, both aggregation methods achieve near-constant coverage across different $k$ values. However, when using Algorithm \ref{algo:randomPermutation}, the EAR remains constant for any $k$, while the EAR for majority vote tends to fluctuate when $k$ is small and increases when $k$ is large. This indicates the performance of majority vote is highly vulnerable to noisy sBPP which has a lower AUC score during the evaluation. In contrast, random permutation is more robust against noisy sBPP, and the choice of $k$ does not significantly affect the final performance. 

{
\paragraph{\textbf{User Study: Evaluating the RTS Questions}}
\begin{table}
\centering
\caption{Schema Linking Performance by Expertise Level}
\begin{small}
\begin{tabular}{lcc}
\toprule
Participant Group& Type   &EM (\%)\\
\midrule
\multirow{2}{*}{Beginner} & Table  & 96.2\\
  & Column & 93.3 \\
\multirow{2}{*}{Expert} & Table  & 98.3\\
  & Column & 95.8\\
\bottomrule
\end{tabular}
\end{small}
\label{User Study: Schema}
\vspace{-1em}
\end{table}
Since RTS relies on user interaction to provide feedback and enhance its reliability, we conducted a user study using the BIRD benchmark to assess how end users with different backgrounds perceive the questions generated by the RTS system. In this task, participants were asked to determine whether a given table or column was relevant to a natural language query. We sampled 100 questions across three difficulty levels as defined by the benchmark: simple, moderate, and challenging. Participants were recruited from varying levels of expertise in data management: the beginner group consisted of undergraduate students with no prior SQL experience, while the expert group comprised students with SQL proficiency. Each group consisted of 10 participants. We recorded participants' accuracy in answering the questions and evaluated their overall schema-linking performance. 
\begin{table*}
\centering 
\caption{Performance on answering RTS generated question by Expertise Level and Query Difficulty} \begin{tabular}{@{}lcccccc@{}} 
\toprule \textbf{Participant Group} & \multicolumn{3}{c}{\textbf{Table Accuracy (\%)}} & \multicolumn{3}{c}{\textbf{Column Accuracy (\%)}} \\ 
\cmidrule(lr){2-4} \cmidrule(lr){5-7} & Simple & Moderate & Challenging & Simple & Moderate & Challenging \\ 
\midrule \textbf{Beginner} & 100 & 96 & 93 & 100 & 92 & 89 \\ 
\textbf{Expert} & 100 & 100 & 99 & 100 & 97 & 94 \\ 
\bottomrule
\end{tabular} 
\label{User Study: QA}
\vspace{-1em}
\end{table*}
{ Table \ref{User Study: Schema} shows the final schema linking performance, while Table \ref{User Study: QA} presents the results of participants answering RTS-generated questions to determine whether a table/column is relevant to the given question. }
The results indicate that both beginner and expert participants were able to assist the RTS system by accurately answering its generated questions, particularly for simple and moderate ones. The performance in schema linking across these categories suggests that the RTS system effectively produces questions that participants can understand, regardless of their expertise.
However, as the difficulty of the questions increased, a gap in answer accuracy between the beginner and expert groups became apparent for both table and column identification, as shown in Table \ref{User Study: QA}. While the expert group maintained high accuracy, the beginner group experienced a more significant drop in performance, particularly in column matching, where accuracy fell to 89\%.

This decline can be attributed to two main factors: first, the relevant part of the schema in the more difficult questions is often complex, with a significantly higher number of attributes; second, currently the questions present the schema in a DDL format, which may not be as easily interpreted by beginners. Those two factors increase the difficulty for the beginner group to fully understand the schema and the relation between each table/column.
Additionally, the schema in challenging questions often requires more detailed descriptions, particularly explaining column names. For example, in a table containing laboratory test results for various biomarkers, it may be difficult to map the abbreviated column name "T-BIL" to a question about total blood bilirubin levels without incorporating external knowledge and/or more explicit labelling.

These learnings motivate us to suggest that future work on RTS may focus on user interaction via the development of a user-friendly HCI framework to improve question interpretability. This may encompass providing a suitable context (possibly enhanced using the LLM itself) along with visually appealing and intuitive layouts of tables and columns to aid question answering. This presents an opportunity for future work in this area.
}
\section{Related Work} \label{sect:related}
\subsection{Uncertainty Quantification}

LLMs have been known to exhibit over-confidence \cite{xiong2024can}, poorly calibrated \cite{jiang2021can, desai2020calibration}, and hallucinations \cite{ji2023survey}, 
leaving the uncertainty quantification a critical problem for real-world applications.
While uncertainty quantification has also been actively studied \cite{abdar2021review, ovadia2019can, gawlikowski2023survey} in deep learning, with typical methods including Bayesian methods \cite{gal2016dropout} and ensemble methods \cite{lakshminarayanan2017simple}, such methods are not suitable for LLM inference due to unique characteristics of LLM such as large space and similar semantics of the outputs \cite{geng2024survey, gawlikowski2023survey}.
Uncertainty quantification for LLM can be divided into two categories: transparent-boxes and opaque-boxes approaches.
Transparent-box approaches utilize additional information from within the model, such as logits \cite{murray2018correcting, vazhentsev2023efficient}, internal states \cite{li2023inference, ren2023out, azaria2023internal}, or semantics \cite{kuhn2023semantic} to quantify uncertainty. However, these approaches require access to internal information and often necessitate additional supervised training \cite{geng2024survey}. Opaque-boxes methods, such as linguistic confidence methods \cite{mielke2022reducing, linteaching} are able to output uncertainty in natural language but require additional calibration procedures \cite{mielke2022reducing, xiong2024can}, while consistency methods \cite{wang2023self, si2023prompting, manakul2023selfcheckgpt} are effective but come with significant computational overhead. Despite the advancements in these methods, none provide correctness guarantees.
Conformal prediction \cite{papadopoulos2002inductive, vovk2005algorithmic, angelopoulos2023conformal, shafer2008tutorial}, on the other hand, offers statistical coverage guarantees while requiring minimal assumptions, attracts research interest on various tasks such as classification \cite{romano2020classification}, machine translation \cite{giovannotti2023evaluating}, and open-domain question answering \cite{angelopoulosconformal}, etc. Conformal prediction is particularly appealing as it requires only a relatively small calibration set \cite{angelopoulos2023conformal}, making it both efficient and flexible.

\subsection{Text-to-SQL}

Text-to-SQL has been studied for decades \cite{katsogiannis2023survey, katsogiannis2021deep}. Traditional methods primarily rely on parsing trees and human-crafted rules \cite{saha2016athena}, which support only subsets of natural language \cite{warren1982efficient, popescu2003towards, zelle1996learning} or require user interactions \cite{li2014constructing}, thus limiting their applications. With advances in deep learning \cite{katsogiannis2021deep}, a major line of work utilizes pre-trained language models (PLM), such as BERT \cite{devlin2018bert, yin2020tabert}, Grappa \cite{yugrappa}, T5 \cite{raffel2020exploring}, approaching the problem as a sequence-to-sequence learning task \cite{sutskever2014sequence} using encoder-decoder architectures \cite{vaswani2017attention,dong2016language}. 
To enforce the grammatical rules of SQL, rather than generating SQL tokens directly \cite{dong2016language,zhong2017seq2sql, lin2020bridging, xu2022sead, scholak2021picard}, most works utilize grammar-based \cite{yu2018syntaxsqlnet, dong2018coarse, guo2019towards, bogin2019global, wang2020rat, rubin2021smbop, gan2021natural, cao2021lgesql} or sketch-based approaches \cite{xu2017sqlnet, yu2018typesql, lee2019clause, ma2020mention, choi2021ryansql, li2023resdsql, fu2023catsql} for decoders. 
However, such approaches require extensive training corpora and the model capabilities may be limited by their sizes and architectures.

LLM-based approaches have drawn much attention recently due to the success of LLM \cite{chang2024survey, yang2024harnessing}. These studies primarily focus on in-context learning settings \cite{brown2020language}, utilizing LLMs with prompt engineering techniques \cite{liu2023pre, gao2024text, li2024can, tai2023exploring}. 
While there is some exploration of prompt design in the zero-shot setting \cite{chang2023prompt}, most of these works concentrate on few-shot methods \cite{gao2024text}, where a limited number of demonstrations are available in the context as examples.
These in-context learning approaches address text-to-SQL from one or more aspects, including question decomposition \cite{pourreza2024din,tai2023exploring, nan2023enhancing,gao2024text,xie2024decomposition,fan2024metasql,ren2024purple, arora2023adapt}, demonstration selection \cite{guo2023retrieval,guo2023prompting,nan2023enhancing,gao2024text,chang2023selective, ren2024purple}, prompt design \cite{tai2023exploring,zhang2023act,guo2023prompting,jiang2023structgpt,gao2024text,hong2024knowledge, arora2023adapt}, and additional procedures such as ranking \cite{fan2024metasql,zhang2023coder,zeng2023n}, self-correction \cite{pourreza2024din}, majority voting \cite{nan2023enhancing}, or utilizing execution results \cite{ni2023lever,chen2024teaching,hong2024knowledge,shi2022natural,guo2023prompting,guo2023retrieval}. 
For instance, DIN-SQL \cite{pourreza2024din} proposes to decompose the problem into subtasks, including schema linking, query classification \& decomposition, SQL generation, and self-correction, while PURPLE \cite{ren2024purple} proposes to focus on demonstration selection with four modules consists of schema pruning, skeleton prediction, demonstration selection, and database adaption.
Additionally, other works have explored applying supervised fine-tuning techniques on LLMs, achieving significant generation speedup \cite{kou2024cllms} and quality improvements \cite{gao2024text,li2024codes}.

\subsection{Schema Linking}

Schema linking \cite{lei2020re,katsogiannis2023survey} has been widely applied due to its pivotal role in improving SQL generation performance \cite{gan2021natural,guo2019towards,wang2020rat} and domain generalization \cite{wang2021meta,lei2020re}. 
Early works utilize simple heuristics such as string matching to identify columns/tables from natural language \cite{guo2019towards,yu2018typesql,brunner2021valuenet}, mainly as a pre-processing step, which could result in inaccurate linking \cite{dong2019data}. To enhance performance, learned methods have been introduced to resolve schema linking either as a separate problem \cite{dong2019data, chang2020zero,ma2020mention}, or as a component in the network, leveraging techniques such as co-attention \cite{zhang2019editing} and graph encoding \cite{bogin2019representing, wang2020rat,cao2021lgesql,li2023graphix,hui2022s2sql,cao2021lgesql,cai2021sadga,wang2022proton, bogin2019global}. 
Recent efforts on LLM-based text-to-SQL also incorporate schema linking as a component to include only related database elements \cite{pourreza2024din}, or to prune unrelated ones based on schema ranking \cite{ren2024purple} for performance improvement. 

\section{Conclusions} \label{sect:conclude}

In this paper, we present the Reliable Text-to-SQL (RTS) framework, which focuses on the schema linking phase and autonomously detects potential errors. We propose the Branching Point Predictor (BPP) that utilizes conformal prediction techniques on the hidden layers of LLM for schema linking with probabilistic guarantees. 
Through extensive experiments on commonly used benchmarks, we validate the effectiveness of our proposed framework, stressing the potential of combining LLMs with human feedback for robust and reliable applications.

\section*{Acknowledgment}

This work was supported in part by the NSERC Discovery Grants. The authors would like to thank the anonymous reviewers, whose valuable comments helped improve
this paper.
\newpage
\bibliographystyle{ACM-Reference-Format}
\bibliography{refer}

\end{document}